\documentclass[letterpaper, paper,11pt]{AAS}		

\usepackage{titlesec}
\titlespacing*{\section}{0pt}{0.5em}{0.5em}
\titlespacing*{\subsection}{0pt}{0.5em}{0.5em}
\usepackage{bm}
\usepackage{amsmath}
\usepackage{subfigure}
\usepackage[colorlinks=true, pdfstartview=FitV, linkcolor=black, citecolor= black, urlcolor= black]{hyperref}
\usepackage{overcite}
\usepackage{footnpag}			      	
\usepackage{amsfonts}
\usepackage{amssymb}
\usepackage{ragged2e}
\usepackage{rotating}
\usepackage[dvipsnames]{xcolor}
\usepackage{algorithm}

\usepackage{acronym}
\acrodef{ko}[KO]{Koopman operator}
\acrodef{dmd}[DMD]{Dynamic Mode Decomposition}
\acrodef{cr3bp}[CR3BP]{Circular Restricted Three Body Problem}
\acrodef{er3bp}[ER3BP]{Elliptical Restricted Three Body Problem}
\acrodef{naff}[NAFF]{Numerical Analysis of Fundamental Frequencies}
\acrodef{fma}[FMA]{Frequency Map Analysis}

\PaperNumber{24-293}

\usepackage{comment}

\begin{document}
\newcommand{\tbf}[1]{\textbf{#1}}

\newcommand{\mbf}[1]{\mathbf{#1}}
\newcommand{\bsym}[1]{\boldsymbol{#1}}
\newcommand{\mcal}[1]{\mathcal{#1}}

\newcommand{\red}[1]{\textcolor{red}{#1}}
\newcommand{\blue}[1]{\textcolor{blue}{#1}}
\newcommand{\green}[1]{\textcolor{green}{#1}}
\newcommand{\vr}[1]{\textbf{\textcolor{VioletRed}{#1}}}
\newcommand{\bo}[1]{\textbf{\textcolor{BurntOrange}{#1}}}
\newcommand{\mb}{\mathbf}
\newcommand{\mc}{\mathcal}
\newcommand{\ba}{\mathbf{a}}
\newcommand{\bc}{\mathbf{c}}
\newcommand{\bdyn}{\mathbf{f}}
\newcommand{\bh}{\mathbf{h}}
\newcommand{\bp}{\mathbf{p}}
\newcommand{\bq}{\mathbf{q}}
\newcommand{\br}{\mathbf{r}}
\newcommand{\bt}{\mathbf{t}}
\newcommand{\bu}{\mathbf{u}}
\newcommand{\bv}{\mathbf{v}}
\newcommand{\bw}{\mathbf{w}}
\newcommand{\bx}{\mathbf{x}}
\newcommand{\by}{\mathbf{y}}
\newcommand{\bz}{\mathbf{z}}
\newcommand{\bA}{\mathbf{A}}
\newcommand{\bB}{\mathbf{B}}
\newcommand{\bC}{\mathbf{C}}
\newcommand{\bD}{\mathbf{D}}
\newcommand{\bE}{\mathbf{E}}
\newcommand{\bG}{\mathbf{G}}
\newcommand{\bH}{\mathbf{H}}
\newcommand{\bL}{\mathbf{L}}
\newcommand{\bM}{\mathbf{M}}
\newcommand{\bP}{\mathbf{P}}
\newcommand{\bQ}{\mathbf{Q}}
\newcommand{\bU}{\mathbf{U}}
\newcommand{\bV}{\mathbf{V}}
\newcommand{\bX}{\mathbf{X}}
\newcommand{\bY}{\mathbf{Y}}
\newcommand{\bsig}{\boldsymbol{\sigma}}
\newcommand{\bom}{\boldsymbol{\omega}}
\newcommand{\bbet}{\boldsymbol{\beta}}
\newcommand{\bs}{\boldsymbol}
\newcommand{\mbb}{\mathbb}
\newcommand{\mbbE}{\mathbb{E}}
\newcommand{\mds}{\mathds}
\newcommand{\mdsone}{\mathds{1}}

\newcommand{\dbp}{\dot{\mathbf{p}}}
\newcommand{\dbr}{\dot{\mathbf{r}}}
\newcommand{\dbv}{\dot{\mathbf{v}}}
\newcommand{\dbx}{\dot{\mathbf{x}}}
\newcommand{\dbH}{\dot{\mathbf{H}}}
\newcommand{\dbR}{\dot{\mathbf{R}}}
\newcommand{\dbV}{\dot{\mathbf{V}}}
\newcommand{\dbsig}{\dot{\boldsymbol{\sigma}}}
\newcommand{\dbom}{\dot{\boldsymbol{\omega}}}

\newcommand{\pd}{\partial}
\newcommand{\pdt}{\partial t}

\newcommand{\dr}{\dot{r}}
\newcommand{\dP}{\dot{P}}
\newcommand{\dQ}{\dot{Q}}
\newcommand{\dR}{\dot{R}}
\newcommand{\dx}{\dot{x}}
\newcommand{\dy}{\dot{y}}
\newcommand{\dv}{\dot{v}}
\newcommand{\dU}{\dot{U}}
\newcommand{\dV}{\dot{V}}
\newcommand{\dW}{\dot{W}}
\newcommand{\dX}{\dot{X}}
\newcommand{\dY}{\dot{Y}}
\newcommand{\dZ}{\dot{Z}}
\newcommand{\dal}{\dot{\alpha}}
\newcommand{\dth}{\dot{\theta}}
\newcommand{\dTh}{\dot{\Theta}}
\newcommand{\ddel}{\dot{\delta}}
\newcommand{\dbTh}{\dot{\boldsymbol{\Theta}}}
\newcommand{\dph}{\dot{\phi}}
\newcommand{\dPh}{\dot{\Phi}}
\newcommand{\dps}{\dot{\psi}}
\newcommand{\dPs}{\dot{\Psi}}
\newcommand{\tps}{\tilde{\psi}}
\newcommand{\tth}{\tilde{\theta}}
\newcommand{\tph}{\tilde{\phi}}
\newcommand{\dOm}{\dot{\Omega}}
\newcommand{\dom}{\dot{\omega}}

\newcommand{\ddbr}{\ddot{\mathbf{r}}}
\newcommand{\ddr}{\ddot{r}}
\newcommand{\ddx}{\ddot{x}}
\newcommand{\ddy}{\ddot{y}}
\newcommand{\ddbR}{\ddot{\mathbf{R}}}
\newcommand{\ddbsig}{\ddot{\boldsymbol{\sigma}}}
\newcommand{\ddth}{\ddot{\theta}}

\newcommand{\cA}{\mathcal{A}}
\newcommand{\cBp}{\mathcal{B}^*}
\newcommand{\cC}{\mathcal{C}}
\newcommand{\cD}{\mathcal{D}}
\newcommand{\calE}{\mathcal{E}}
\newcommand{\cF}{\mathcal{F}}
\newcommand{\cG}{\mathcal{G}}
\newcommand{\cH}{\mathcal{H}}
\newcommand{\calI}{\mathcal{I}}
\newcommand{\calJ}{\mathcal{J}}
\newcommand{\cK}{\mathcal{K}}
\newcommand{\cN}{\mathcal{N}}
\newcommand{\cO}{\mathcal{O}}
\newcommand{\cT}{\mathcal{T}}
\newcommand{\cW}{\mathcal{W}}
\newcommand{\cBI}{\mathcal{BI}}
\newcommand{\cIB}{\mathcal{IB}}

\newcommand{\sA}{\mathscr{A}}
\newcommand{\sB}{\mathscr{B}}
\newcommand{\sC}{\mathscr{C}}
\newcommand{\sD}{\mathscr{D}}
\newcommand{\sE}{\mathscr{E}}
\newcommand{\sF}{\mathscr{F}}

\newcommand{\mbA}{\mathbb{A}}
\newcommand{\mbD}{\mathbb{D}}
\newcommand{\mbF}{\mathbb{F}}
\newcommand{\mbW}{\mathbb{W}}

\newcommand{\mtT}{\mathtt{T}}

\newcommand{\uer}{\hat{\mathbf{e}}_r}  
\newcommand{\uef}{\hat{\mathbf{e}}_f}  
\newcommand{\ueR}{\hat{\mathbf{e}}_R}  
\newcommand{\ueth}{\hat{\mathbf{e}}_\theta}
\newcommand{\uiye}{\hat{\mathbf{i}}_y}
\newcommand{\ueze}{\hat{\mathbf{e}}_z}
\newcommand{\ueph}{\hat{\mathbf{e}}_\phi}
\newcommand{\ui}{\hat{\mathbf{i}}}  
\newcommand{\uj}{\hat{\mathbf{j}}}  
\newcommand{\uk}{\hat{\mathbf{k}}}  
\newcommand{\ue}{\hat{\mathbf{e}}} 
\newcommand{\ub}{\hat{\mathbf{b}}} 
\newcommand{\ud}{\hat{\mathbf{d}}}
\newcommand{\uR}{\hat{\mathbf{R}}} 
\newcommand{\us}{\hat{\mathbf{s}}}  
\newcommand{\ut}{\hat{\mathbf{t}}} 
\newcommand{\un}{\hat{\mathbf{n}}} 
\newcommand{\uv}{\hat{\mathbf{v}}}
\newcommand{\um}{\hat{\mathbf{m}}}  
\newcommand{\uix}{\hat{\mathbf{i}}_1}  
\newcommand{\uiy}{\hat{\mathbf{i}}_2}  
\newcommand{\uiz}{\hat{\mathbf{i}}_3}    
\newcommand{\uie}{\hat{\mathbf{i}}_e}  
\newcommand{\uih}{\hat{\mathbf{i}}_h}  
\newcommand{\ubx}{\hat{\mathbf{b}}_1}  
\newcommand{\uby}{\hat{\mathbf{b}}_2}  
\newcommand{\ubz}{\hat{\mathbf{b}}_3} 
\newcommand{\ubj}{\hat{\mathbf{b}_j}}     
\newcommand{\uex}{\hat{\mathbf{e}}_1}  
\newcommand{\uey}{\hat{\mathbf{e}}_2}  
\newcommand{\uez}{\hat{\mathbf{e}}_3}    
\newcommand{\uax}{\hat{\mathbf{a}}_1}  
\newcommand{\uay}{\hat{\mathbf{a}}_2}  
\newcommand{\uaz}{\hat{\mathbf{a}}_3}  

\newcommand{\bam}{\mathbf{R} \times \dot{\mathbf{R}}}	
\newcommand{\ftbp}{-\frac{\mu}{R^3} \mathbf{R}}
\newcommand{\sumin}{\sum_{i=1}^N}
\newcommand{\sumjn}{\sum_{j=1}^N}
\newcommand{\sumkn}{\sum_{k=1}^N}
\newcommand{\sumtt}{\sum_{t \; \in \; T}}
\newcommand{\sP}{\; + \;}
\newcommand{\sM}{\; - \;}
\newcommand{\rbbast}{{\color{red}\boldsymbol{\bigoast}}}
\newcommand{\seq}{\; = \;}
\newcommand{\sdef}{\; \triangleq \;}
\newcommand{\sleq}{\; \leq \;}
\newcommand{\sgeq}{\; \geq \;}
\newcommand{\sfor}{\; \forall \;}
\newcommand{\such}{\; | \;}
\newcommand{\ssin}{\; \in \;}
\newcommand{\sfl}{\forall \;}
\newcommand{\unit}{\bigcup_{t \in T}}
\newcommand{\intt}{\bigcap_{t \in T}}
\newcommand{\unii}{\bigcup_{i \in I}}
\newcommand{\inti}{\bigcap_{i \in I}}
\newcommand{\sse}{\subseteq}
\newcommand{\salg}{\sigma\text{-algebra}}
\newcommand{\ph}{\frac{\Phi}{2}}
\newcommand{\cpa}{\cos \Phi}
\newcommand{\spa}{\sin \Phi}
\newcommand{\cpah}{\cos \frac{\Phi}{2}}
\newcommand{\spah}{\sin \frac{\Phi}{2}}
\newcommand{\ombi}{\boldsymbol{\omega}^{\mathcal{B}/\mathcal{I}}}
\newcommand{\ombib}{\boldsymbol{\omega}^{\mathcal{B}/\mathcal{I}}_\mathcal{B}}
\newcommand{\ombii}{\boldsymbol{\omega}^{\mathcal{B}/\mathcal{I}}_\mathcal{I}}
\newcommand{\rbi}{\mathbf{R}_\mathcal{BI}}
\newcommand{\rba}{\mathbf{R}_\mathcal{BA}}
\newcommand{\rbit}{\tilde{\mathbf{R}}_\mathcal{BI}}
\newcommand{\rib}{\mathbf{R}_\mathcal{IB}}
\newcommand{\rab}{\mathbf{R}_\mathcal{AB}}
\newcommand{\ombiic}{\left[\boldsymbol{\omega}^{\mathcal{B}/\mathcal{I}}_\mathcal{I}\right]^\times}
\newcommand{\ombibc}{\left[\boldsymbol{\omega}^{\mathcal{B}/\mathcal{I}}_\mathcal{B}\right]^\times}
\newcommand{\tom}{\tilde{\omega}}
\newcommand{\wex}{\widetilde{\exp}}
\newcommand{\com}{\dbinom}
\newcommand{\vinf}{V_\infty}
\newcommand{\clz}{C_{L_0}}
\newcommand{\clal}{C_{L_{\alpha}}}
\newcommand{\clalw}{C_{L_{\alpha_w}}}
\newcommand{\clalh}{C_{L_{\alpha_H}}}
\newcommand{\cldal}{C_{L_{\dot{\alpha}}}}
\newcommand{\clq}{C_{L_Q}}
\newcommand{\clih}{C_{L_{i_H}}}
\newcommand{\clde}{C_{L_{\delta E}}}
\newcommand{\cybe}{C_{Y_\beta}}
\newcommand{\cyp}{C_{Y_P}}
\newcommand{\cyr}{C_{Y_R}}
\newcommand{\cyda}{C_{Y_{\delta A}}}
\newcommand{\cydr}{C_{Y_{\delta R}}}
\newcommand{\cdz}{C_{D_0}}
\newcommand{\cdal}{C_{D_\alpha}}
\newcommand{\cdde}{C_{D_{\delta E}}}
\newcommand{\clbe}{C_{L_\beta}}
\newcommand{\clp}{C_{L_P}}
\newcommand{\clr}{C_{L_R}}
\newcommand{\clda}{C_{L_{\delta A}}}
\newcommand{\cldr}{C_{L_{\delta R}}}
\newcommand{\cmz}{C_{M_0}}
\newcommand{\cmal}{C_{M_\alpha}}
\newcommand{\cmdal}{C_{M_{\dot{\alpha}}}}
\newcommand{\cmq}{C_{M_Q}}
\newcommand{\cmih}{C_{M_{i_H}}}
\newcommand{\cmde}{C_{M_{\delta E}}}
\newcommand{\cnbe}{C_{N_\beta}}
\newcommand{\cnr}{C_{N_R}}
\newcommand{\cnp}{C_{N_P}}
\newcommand{\cnda}{C_{N_{\delta A}}}
\newcommand{\cndr}{C_{N_{\delta R}}}
\newcommand{\scoll}{\{A_t \}_{t \in T}}
\newcommand{\bPlong}{\mathbf{P}_\text{long}}
\newcommand{\bQlong}{\mathbf{Q}_\text{long}}
\newcommand{\bTlong}{\mathbf{T}_\text{long}}
\newcommand{\bPlat}{\mathbf{P}_\text{lat}}
\newcommand{\bQlat}{\mathbf{Q}_\text{lat}}
\newcommand{\bTlat}{\mathbf{T}_\text{lat}}
\newcommand{\vless}{\preccurlyeq}
\newcommand{\vmore}{\succcurlyeq}

\newcommand{\tor}{\mathbb{T}_0}
\newcommand{\rvb}{\mathbb{b}}
\newcommand{\rvu}{\mathbb{U}}
\newcommand{\rvv}{\mathbb{V}}
\newcommand{\rvw}{\mathbb{W}}
\newcommand{\rv}{\mathbb{X}}
\newcommand{\rvy}{\mathbb{Y}}
\newcommand{\rvz}{\mathbb{Z}}
\newcommand{\brv}{\mathds{X}}
\newcommand{\rvi}{\mathbb{X}^{-1}}
\newcommand{\rvsq}{\{\mathbb{X}_n\}}
\newcommand{\circm}{\textcircled{m}}
\newcommand{\expec}{\mathbb{E}}
\newcommand{\pspace}{(\Omega, \mathscr{F}, P)}
\newcommand{\PDF}{F_\mathbb{X}}
\newcommand{\PDFY}{F_\mathbb{Y}}
\newcommand{\PDFZ}{F_\mathbb{Z}}
\newcommand{\JPDF}{F_\mathds{X}}
\newcommand{\PDFXY}{F_\mathbb{XY}}
\newcommand{\PDFXCY}{F_{\mathbb{X}|\mathbb{Y}}}
\newcommand{\PDFYCX}{F_{\mathbb{Y}|\mathbb{X}}}
\newcommand{\pdfto}{f_{\mathbb{T}_0}}
\newcommand{\pdf}{f_\mathbb{X}}
\newcommand{\pdfy}{f_\mathbb{Y}}
\newcommand{\pdfz}{f_\mathbb{Z}}
\newcommand{\jpdf}{f_\mathds{X}}
\newcommand{\pdfxy}{f_{\mathbb{XY}}}
\newcommand{\pdfxcy}{f_{\mathbb{X}|\mathbb{Y}}}
\newcommand{\pdfycx}{f_{\mathbb{Y}|\mathbb{X}}}
\newcommand{\pgfto}{G_{\mathbb{T}_0}}
\newcommand{\pgf}{G_\mathbb{X}}
\newcommand{\pgfy}{G_\mathbb{Y}}
\newcommand{\pgfz}{G_\mathbb{Z}}
\newcommand{\mgf}{M_\mathbb{X}}
\newcommand{\mgfy}{M_\mathbb{Y}}
\newcommand{\mgfz}{M_\mathbb{Z}}
\newcommand{\mgfxy}{M_\mathbb{XY}}
\newcommand{\cf}{\phi_\mathbb{X}}
\newcommand{\cfy}{\phi_\mathbb{Y}}
\newcommand{\cfz}{\phi_\mathbb{Z}}
\newcommand{\cfxy}{\phi_\mathbb{XY}}
\newcommand{\rvom}{\mathbb{X}(\omega)}
\newcommand{\mixo}{(-\infty,\, x]}
\newcommand{\miyo}{(-\infty,\, y]}
\newcommand{\oO}{\{ \omega \in \Omega }
\newcommand{\vrv}{(\mathbb{X}_1, \mathbb{X}_2, \ldots, \mathbb{X}_N)'}
\newcommand{\rvmu}{\langle \, \mathbb{X} \, \rangle}
\newcommand{\nog}{\frac{1}{\sqrt{2\pi \sigma^2}}}
\newcommand{\nobg}{\frac{1}{(2 \pi)^{N/2}\sqrt{\text{det}(\mathbf{P})}}}
\newcommand{\gauss}{\exp\left(- \frac{(x-\mu)^2}{2\sigma^2} \right)}
\newcommand{\bgauss}{\exp\left(-\frac{1}{2}(\mathbf{x}-\boldsymbol{\mu})^\text{T} \mathbf{P}^{-1}(\mathbf{x}-\boldsymbol{\mu}) \right)}
\newcommand{\toas}{\xrightarrow{\text{a.s.}}}
\newcommand{\toip}{\xrightarrow{\text{i.p.}}}
\newcommand{\topr}{\xrightarrow{\text{P}}}
\newcommand{\tolaw}{\xrightarrow{\text{law}}}
\newcommand{\tod}{\xrightarrow{\text{D}}}
\newcommand{\tof}{\xrightarrow{F}}
\newcommand{\todel}{\xrightarrow{\Delta}}
\newcommand{\tolt}{\xrightarrow{L_2}}
\newcommand{\tolp}{\xrightarrow{L_p}}
\newcommand{\spx}{\mathbb{X}_t}
\newcommand{\spy}{\mathbb{Y}_t}
\newcommand{\spz}{\mathbb{Z}_t}
\newcommand{\spxc}{\{\mathbb{X}_t\}_{t \in T}}
\newcommand{\mux}{\mu_\mathbb{X}}
\newcommand{\kxx}{K_\mathbb{XX}}
\newcommand{\rxx}{R_\mathbb{XX}}
\newcommand{\rxy}{R_\mathbb{XY}}
\newcommand{\xseq}{\{x_n\}}
\newcommand{\yseq}{\{y_n\}}

\newcommand{\limni}{\lim_{n \rightarrow \infty}}
\newcommand{\limnmi}{\lim_{n \rightarrow -\infty}}
\newcommand{\limii}{\lim_{i \rightarrow \infty}}
\newcommand{\limimi}{\lim_{i \rightarrow -\infty}}
\newcommand{\intii}{\int_{-\infty}^\infty}
\newcommand{\intoi}{\int_{0}^\infty}
\newcommand{\intix}{\int_{-\infty}^x}

\newcommand{\bde}{\begin{defin}}
\newcommand{\ede}{\end{defin}}
\newcommand{\bit}{\begin{itemize}}
\newcommand{\eit}{\end{itemize}}
\newcommand{\bse}{\begin{subequations}}
\newcommand{\ese}{\end{subequations}}
\newcommand{\beq}{\begin{equation}}
\newcommand{\eeq}{\end{equation}}
\newcommand{\bal}{\begin{align}}
\newcommand{\eal}{\end{align}}
\newcommand{\bcas}{\begin{cases}}
\newcommand{\ecas}{\end{cases}}
\newcommand{\non}{\nonumber}
\newcommand{\uns}{\underset}
\newcommand{\unb}{\underbrace}
\newcommand{\lcb}{\left(}
\newcommand{\rcb}{\right)}
\newcommand{\lccb}{\left\{}
\newcommand{\rccb}{\right\}}
\newcommand{\lsb}{\left[}
\newcommand{\rsb}{\right]}
\newcommand{\bBm}{\begin{Bmatrix}}
\newcommand{\eBm}{\end{Bmatrix}}
\newcommand{\ovl}{\overline}
\newcommand{\tb}{\textbf}
\newcommand{\tit}{\textit}
\newcommand{\ttt}{\texttt}
\newcommand{\tup}{\textup}
\newcommand{\tx}{\text}
\newcommand{\txT}{\text{T}}
\newcommand{\txeq}{\text{Eq.}}
\newcommand{\txeqs}{\text{Eqs.}}
\newcommand{\nod}{\noindent}
\newcommand{\examp}{\nod \ding{163} \tb{Example } }
\newcommand{\clrr}{\color{red}}
\newcommand{\clrb}{\color{blue}}
\newcommand{\hs}{\hspace{2mm}}
\newcommand{\vs}{\vspace{2mm}}

\newcommand{\lins}{\mathbf{A} \mathbf{x} \; + \; \mathbf{B} \mathbf{u}}
\newcommand{\rt}{x^*(t)}
\newcommand{\yol}{y_{\text{ol}}}
\newcommand{\ycl}{y_{\text{cl}}}
\newcommand{\brt}{\mathbf{x}^*(t)}
\newcommand{\ssrt}{x^*_{\text{ss}}(t)}
\newcommand{\ssr}{x^*_\text{ss}}
\newcommand{\bref}{\mathbf{x}^*}
\newcommand{\bsref}{\mathbf{x}^*_{\text{ss}}}
\newcommand{\dbssr}{\dot{\mathbf{x}}^*_{\text{ss}}(t)}
\newcommand{\bpert}{\boldsymbol{\delta}}
\newcommand{\linsys}{\mathbf{A x} + \mathbf{B u} }
\newcommand{\ilt}{\mathcal{L}^{-1}}
\newcommand{\sstf}{\mathbf{C}(s \mathbf{I} - \mathbf{A})^{-1} \mathbf{B} + \mathbf{D} }
\newcommand{\iltf}{\mathcal{L}^{-1} \{ F(s) \}}
\newcommand{\iltx}{\mathcal{L}^{-1} \{ X(s) \}}
\newcommand{\ilty}{\mathcal{L}^{-1} \{ Y(s) \}}
\newcommand{\ltf}{\mathcal{L}\{ f(t) \}}
\newcommand{\ltg}{\mathcal{L}\{ g(t) \}}
\newcommand{\ltr}{\mathcal{L}\{ r(t) \}}
\newcommand{\ltu}{\mathcal{L}\{ u(t) \}}
\newcommand{\ltx}{\mathcal{L}\{ x(t) \}}
\newcommand{\lty}{\mathcal{L}\{ y(t) \}}
\newcommand{\tfux}{\frac{X(s)}{U(s)}}
\newcommand{\tfuy}{\frac{Y(s)}{U(s)}}
\newcommand{\tfrx}{\frac{X(s)}{R(s)}}
\newcommand{\tfry}{\frac{Y(s)}{R(s)}}
\newcommand{\tfxy}{\frac{Y(s)}{X(s)}}
\newcommand{\Gp}{G_{\text{plant}}(s)}
\newcommand{\Gc}{G_{\text{control}}(s)}
\newcommand{\Gs}{G_{\text{sensor}}(s)}
\newcommand{\Gsys}{G_{\text{sys}}(s)}
\newcommand{\Gcl}{G_{\text{cl}}(s)}
\newcommand{\Gol}{G_{\text{ol}}(s)}
\newcommand{\stse}{\texttt{SSE}}
\newcommand{\po}{\texttt{PO}}
\newcommand{\ssos}{\texttt{SSOS}}
\newcommand{\pid}{\texttt{PID}}
\newcommand{\gjw}{G(j \omega)}
\newcommand{\gmag}{|G(j \omega)|}
\newcommand{\gphs}{\angle G(j \omega)}
\newcommand{\gmagdb}{|G(j \omega)|_{\texttt{dB}}}
\newcommand{\gmagi}{|G(j \omega_i)|}
\newcommand{\gphsi}{\angle G(j \omega_i)}
\newcommand{\gmagidb}{|G(j \omega_i)|_{\texttt{dB}}}
\newcommand{\ysss}{y_{sss}(t)}
\newcommand{\db}{\texttt{dB}}
\newcommand{\logt}{\log_{10}}
\newcommand{\Delol}{\Delta_\tx{OL}(s)}
\newcommand{\Delcl}{\Delta_\tx{CL}(s)}

\newcommand{\matlab}{MATLAB$^\copyright$}
\newcommand{\bode}{\texttt{bode}}
\newcommand{\lsim}{\texttt{lsim}}
\newcommand{\dmat}{\texttt{.mat}}
\newcommand{\rlocus}{\texttt{rlocus}}
\newcommand{\step}{\texttt{step}}
\newcommand{\impulse}{\texttt{impulse}}
\newcommand{\tf}{\texttt{tf}}

\title{Time-delayed Dynamic Mode Decomposition for families of periodic trajectories in Cislunar Space}

\author{Sriram Narayanan\thanks{Department of Mechanical and Aerospace Engineering,
The Ohio State University,
201 W 19th Ave,
Columbus, 43210, Ohio, USA},  
Mohamed Naveed Gul Mohamed\thanks{Department of Aerospace Engineering,
Texas A \& M University,
710 Ross St,
College Station, 77843, Texas, USA},
Indranil Nayak\thanks{Department of Electrical and Computer Engineering,
The Ohio State University,
2015 Neil Ave,
Columbus, 43210, Ohio, USA}, 
Suman Chakravorty\footnotemark[2],
\ and Mrinal Kumar\footnotemark[1]
}

\maketitle{}

\begin{abstract}
In recent years, the development of the Lunar Gateway and Artemis missions has renewed interest in lunar exploration, including both manned and unmanned missions. This interest necessitates accurate initial orbit determination (IOD) and orbit prediction (OP) in this domain, which faces significant challenges such as severe nonlinearity, sensitivity to initial conditions, large state-space volume, and sparse, faint, and unreliable measurements. This paper explores the capability of data-driven Koopman operator-based approximations for OP in these scenarios. Three stable periodic trajectories from distinct cislunar families are analyzed. The analysis includes theoretical justification for using a linear time-invariant system as the data-driven surrogate. This theoretical framework is supported by experimental validation. Furthermore, the accuracy is assessed by comparing the spectral content captured to period estimates derived from the fast Fourier transform (FFT) and Poincare-like sections.
\end{abstract}

\section{Introduction}

Given the renewed interest from commercial and government entities, the application of data-driven methods in the cislunar orbital regime is essential for supporting mission development and daily space operations. A fundamental question is: \textit{How accurately does the \ac{cr3bp} represent the Earth-Moon-Spacecraft (SO) three-body problem (3BP) and the perturbed Earth-Moon-SO 3BP?} This model gap may introduce significant discrepancies and hinder accurate modeling in this regime.
Traditionally, astrodynamics has relied on analytical methods and physics-based reference models, such as Keplerian mechanics and the \ac{cr3bp}. These models offer insights strongly correlated with the underlying physics, provide unparalleled visualization of trajectories and orbital regimes, and assist in optimal design. With data-driven models we aim to bridge the gap between reference dynamic models and the "true physical process" (perturbed dynamics) within an explainable framework, such as DMD and its Koopman-based counterparts.
This paper is our first step toward building such augmented models. Here, the data-driven approach is applied to the \ac{cr3bp} and analyzed via the spectral features of the dataset. Future work will extend these techniques to more complex and chaotic scenarios, ultimately addressing the model gap and developing augmented physics-based data-driven models.

This paper extends our prior submission (AAS 24-121) to families of periodic orbits in the cislunar regime. We develop a method based on the Koopman operator to construct surrogate models of objects belonging to specific periodic orbit groups in the cislunar regime and subsequently predict their future states. The Koopman operator embeds nonlinear dynamical systems within a global linear framework, thereby leveraging linear systems theory and corresponding techniques to analyze complex nonlinear systems. Dynamic Mode Decomposition (DMD) and its variants serve as a computationally feasible method that derives an optimal linear finite-dimensional approximation of the infinite-dimensional Koopman operator~\cite{schmid2010dynamic, schmid2022dynamic, kutz2016dynamic}. Note that there is no explicit representation of the Koopman operator. Its behavior is determined through its action on a finite-dimensional observable space. Data-driven techniques like DMD are employed to approximate the Koopman modes and to find the most accurate approximation of the Koopman operator within a finite-dimensional space. The spectral decomposition of the Koopman operator offers valuable insights into the behavior of nonlinear systems by identifying distinct modes and their corresponding frequencies. The accurate depiction of the dynamical system's spectral content by DMD facilitates reliable predictions concerning its spatio-temporal evolution. There is a wide array of applications of the Koopman operator for solving problems in orbital mechanics, ranging from building data-driven surrogates for orbital trajectories in the two-body and CR$3$BP regimes~\cite{linares2019koopman,power23,servadio2023koopman}, to attitude control~\cite{Chen2023} and optimal control problems~\cite{jenson22}. The Hankel variant of DMD is specifically used in this work to incorporate the use of time delays. Suppose the spectral content of the data is captured correctly. In that case, accurate estimates can be made for several periods forward in time, under the assumption that the spectral content is time-invariant. 
In the time-varying scenario, an online approach may be necessary, where the system matrix $\mbf{A}$ is continuously updated as we iterate through accrued observations. In this work, we consider stable periodic trajectories so a time-invariant approximation suffices. 
The real benefit of the DMD approach is likely in the xGEO-cislunar regime where orbital elements are difficult to define, but there still exist periodic orbits. A data-driven approach can explicitly identify the spectral content, and when it is invariant, the linear surrogate identified by DMD can predict the motion of the object accurately for several periods forward in time.
This paper is organized as follows. Sec.~\ref{Sec:PF} formulates the problem considered and defines the objective of this work. Sec.~\ref{Sec:MB} provides the necessary mathematical background and discusses the theory behind the use of autoregressive models such as Hankel-DMD to predict the evolution of periodic orbits of nonlinear systems using a finite number of harmonics. Sec.~\ref{Sec:Res} presents observations and results from the conducted experimental analysis.  Finally, Sec.~\ref{Sec:Conc} provides a general discussion summarizing the main contributions of this work and concludes the paper elaborating on planned future work.

\section{Problem Definition}
\label{Sec:PF}
This work focuses on the data-driven modeling and prediction of three cislunar trajectories from families of periodic orbits of significant interest for Earth-Moon missions: an L1-Halo orbit, an L1-L2 butterfly orbit, and a Resonant (2:1) orbit~\cite{folta2004survey, whitley2018earth, frueh2021cislunar, farquhar1971utilization, holzinger2021primer}. In this context, L1 and L2 denote the first two Lagrange points in the Earth-Moon system. Koopman operator-based DMD is the method of choice to derive a data-driven surrogate, predict future states and study properties of the system such as its period and stability. Consider the nonlinear system: $\mathbf{x}(k+1) = \mathbf{f}(\mathbf{x}(k))$ with known initial condition $\mbf{x}(0)$ and some period $\exists \ N > 0$, such that $\mbf{x}(k+N) = \mbf{x}(k) \ \forall \ k $ with unknown system dynamics $\mbf{f}(\cdot)$. Given, a finite time history of the states $\{\mbf{x}(k), \mbf{x}(k+1), \cdots, \mbf{x}(k+l) \}$, this work seeks to find a surrogate model for this nonlinear system via DMD to predict the next state $\mbf{x}(k+l+1)$, and subsequently, forecast the evolution of the system. In addition, the period of this orbit will be found via the fundamental frequency obtained from DMD and the stability properties will be evaluated by studying the eigenvalues of the data-driven surrogate system matrix. In this process, the number of time delays required, the training window size, and how far the predictions can be made using these two model parameters will also be presented.

\section{Mathematical Background}
\label{Sec:MB}
The following section briefly reviews the equations of motion for cislunar dynamics, specifically the \ac{cr3bp} as used in this work. Subsequently discussed, is the theory of the Koopman operator, the data-driven approximation using DMD, and the application of an autoregressive model for predicting periodic trajectories.

\subsection{Cislunar Dynamics}
The motion of the Earth-Moon system and a third body such as a spacecraft is governed by the n-body equations of motion with perturbing forces, where $n=3$. Restricted versions of this problem such as the \ac{cr3bp} and \ac{er3bp} are used in practice to derive analytical methods with considerable structure. This approach, while inaccurate when compared to the full-three-body problem is often adopted because of the existence of families of periodic solutions and the ability to leverage analytical theories on nonlinear dynamics and attractors. 

The \ac{cr3bp} is a simplified version of the general three-body problem used to characterize the motion of a massless particle, such as a spacecraft, in the presence of two massive bodies, such as the Earth and the Moon. The \ac{cr3bp} equations of motion are given in a rotating coordinate frame $\mc{E} = \{ \mbf{e}_x, \mbf{e}_y, \mbf{e}_z \}$ with origin $\mc{O}$ at the mutual barycenter of the two primaries. The $\mbf{e}_x$ vector points toward the moon, $\mbf{e}_z$ is aligned with the angular momentum vector of the system, and $\mbf{e}_y$ points in the direction that completes the dextral orthogonal triad. The \ac{cr3bp} is nondimensionalized such that the distance between the two primaries, the sum of the mass of the two primaries, the rotation rate of the system, and the gravitational parameter are all unity. The constant, $\mu$ relates all normalized measurements and is computed by dividing the smaller primary's mass by the system's total mass. In the absence of control inputs and noise, the equations of motion are given as follows~\cite{szebehely67}, the continuous form is adopted for simplicity,
\begin{equation}
\begin{aligned}
& \Ddot{x}=2 \dot{y}+\frac{\partial U}{\partial x}, \\
& \Ddot{y}=-2 \dot{x}+\frac{\partial U}{\partial y}, \\
& \Ddot{z}=\frac{\partial U}{\partial z},
\end{aligned}
\end{equation}
where the augmented/effective potential $U$ is,
\begin{equation}
U=\frac{1}{2}\left(x^2+y^2\right)+\frac{1-\mu}{\sqrt{(x+\mu)^2+y^2+z^2}}+\frac{\mu}{\sqrt{(x-1+\mu)^2+y^2+z^2}} .
\end{equation}
Note that, the dynamics within the circular restricted three-body system are solely determined by the three-body constant, $\mu$. As $\mu$ approaches zero, the system's dynamics converge towards that of a two-body system, albeit within a rotating reference frame. The \ac{er3bp} uses a non-uniformly rotating coordinate frame with pulsating coordinates and the true anomaly of the second primary as the independent variable to derive the equations of motion, a concise development can be found in Ref~\cite{szebehely64}.
In this work, three stable periodic trajectories in \ac{cr3bp} are considered: an L1-Halo orbit, an L1-L2 butterfly orbit, and a Resonant(2:1) orbit. Koopman operator-based DMD is applied to predict the trajectory and determine properties such as its period and stability. The next section briefly discusses the Koopman operator, DMD—a data-driven method to find approximations of the Koopman operator—and the theory of adopting an autoregressive approach to model periodic trajectories, forming this work's basis.

\subsection{Koopman Operator Theory}

Koopman operator theory allows the study of complex nonlinear dynamics via an infinite dimensional linear operator on the Hilbert space of observable functions~\cite{koopman1932dynamical}. Per Koopman's theory, the state-space corresponding to a nonlinear dynamical system can be transformed into an infinite dimensional observable space where the dynamics is linear. Now, various machine learning approaches, e.g., DMD, Koopman autoencoders (KAE), and others hypothesize that there exists a data transformation under which an approximate finite-dimensional Koopman operator (KO) is available \cite{rowley2009spectral,schmid2022dynamic,azencot2020forecasting}.
Consider the discrete-time deterministic nonlinear system: 
    $\mb{x}(k+1) = \mb{f}(\mb{x}(k))$, 
with known initial condition $\mb{x}(0)$, where $\mb{x}(k) \in \mbb{R}^n$, where the dynamics $\mb{f}(\cdot)$ of the system is unknown. 
The discrete-time KO $\mc{K}$ operates on the space of \tb{observables},
$g(\mb{x})$, such that the evolution of observables is explained by $\cK$ as follows
\begin{equation}
 \mc{K} g\left(\mb{x}(k)\right) \triangleq \left(g \circ \mb{f}\right)\left(\mb{x}(k)\right)=g\left(\mb{x}(k+1)\right)
 \label{Eq:koopman}.
\end{equation}
The KO $\mc{K}$ is linear but infinite-dimensional. The scalar observables $g$ are also referred to as the Koopman observables. The eigendecomposition of the KO gives us ~\cite{mohr2014construction}: 
\begin{align}
    \mc{K}\theta_m(\mb{x}) = \lambda_m \theta_m(\mb{x}), \ \ m = 0,1,2,....,N-1
    \label{eq:koop1},
\end{align} 
where $\theta_m(\mb{x})$ and $\lambda_m$ are corresponding Koopman eigenfunctions and eigenvalues. Assuming eigenfunctions span the observable space, a vector-valued observable $\mb{h}$ is defined in terms of the eigenfunctions,
\begin{align}
    \mb{h}(\mb{x}) =  \sum_{m=0}^{\infty} \theta_m (\mb{x})\bs{\psi}_m, \ \  \bs{\psi}_m \in \mathbb{C}^n
    \label{eq:koop2}.
\end{align}
Substituting Eq.~\eqref{eq:koop1} into Eq.~\eqref{eq:koop2}~~\cite{nayak2021detection}, 
\begin{align}
    \mb{h}(\mb{F}^k(\mb{x}(0))) = 
    \mb{h}(\mb{x}(k)) =  \sum_{m=0}^{\infty} \lambda_m^k \theta_m (\mb{x}(0))\bs{\psi}_m, \ \  \bs{\psi}_m \in \mathbb{C}^n,
    \label{eq:koop3}
\end{align}
where $\bs{\psi}_m$ are the Koopman modes. 
The time expansion of vector-valued observable $\mb{h}$ completely determines the evolution of the nonlinear system from its initial condition. The above developments allow us to capture the evolutionary mechanics of observables, which could be the partially measured state of the dynamics, or other functional forms of the state (e.g., range, bearings, etc.). 
Despite its attractive linear structure, there is no explicit representation of the Koopman operator. Its behavior is inferred through its action on a finite-dimensional observable space. Data-driven techniques, like DMD, are typically employed to approximate the Koopman modes. DMD aims to find the most accurate approximation of $\mc{K}$ within a finite-dimensional space. The spatiotemporal modes derived through DMD exhibit convergence to the Koopman modes when applied to a specific set of linear observables while assuming that state variables themselves serve as the set of observables $\mb{h(x)} = \mb{x}$ ~\cite{rowley2009spectral,tu2013dynamic}. 
The next section discusses the use of an autoregressive approach to model periodic trajectories and in particular the use of the Hankel variant of the DMD algorithm which allows the use of time-delayed embedding in the snapshot matrices to improve spatial resolution.

\subsection{Modeling Periodic Trajectories: Auto-regression and Hankel DMD}
\label{Sec:Th}
The nonlinear system $\mbf{x}(k+1) = \mbf{f}(\mbf{x}
(k))$ is considered, where $\mbf{x}(0)$ is the initial condition and there exists a non-constant periodic solution i.e. $\exists \ N > 0$ such that $\mbf{x}(k+N) = \mbf{x}(k) \ \forall \ k$. Such trajectories may be modeled using a linear autoregressive (AR) approach by using the DMD algorithm. The authors elaborate on this approach in Ref.~\cite{narayanan2024predictive}. A brief development is presented here for completeness. As $\mbf{x}(k)$ is a discrete-time signal with period $N$, its Fourier series can be written as follows \cite{signals_systems}, 
\begin{align}
    \mbf{x}(k) = \sum_{m = 0}^{N-1} \mbf{a}_m e^{j \omega_m k}, \label{Eq:Fourier_series}
\end{align}
where, $\mbf{a}_m \in \mathbb{C}^n$ are Fourier coefficients and $\phi_m(k) = e^{j \omega_m k}$ are the basis functions. And the coefficients $\mbf{a}_m$ can be written in terms of the signal $\mbf{x}(k)$ as 
\begin{align}
    \mbf{a}_m = \sum_{k=0}^{N-1} \mbf{x}(k)e^{-j \omega_m k}.
\end{align}
Per Nyquist's criterion, the Fourier series in Eq.~\eqref{Eq:Fourier_series} can have a maximum of $N/2$ frequencies. 
Since both the negative and the positive components are represented, Eq.~\eqref{Eq:Fourier_series} has $2\times (N/2)$ terms. If the signal $\mbf{x}(k)$ has only $M$ frequencies, where, $1 \leq M \leq N/2 $, then, the Fourier series can be written as 
\begin{align}
    \mbf{x}(k) = \sum_{m = 0}^{2M-1} \mbf{a}_m e^{j \omega_m k}, \label{Eq:Fourier_series_M}
\end{align}
where, $\{\omega_0, \omega_1, \cdots, \omega_{2M-1} \}$ are the different frequencies in the signal. Given the system's response $\{\mbf{x}(k), \mbf{x}(k+1), \cdots \mbf{x}(k+l)\}$, one can estimate the Fourier coefficients by solving the linear least squares problem,
$\bar{\boldsymbol{\alpha}} := \mbf{V}^{\dagger} \mbf{Y}$ where, the matrices $\mbf{V}$ and $\mbf{Y}$ are~\cite{narayanan2024predictive},
\begin{align*}
    \mbf{V} =& 
    \begin{bmatrix}
        e^{j \omega_0} & 0 & 0 & \cdots & 0 \\
        0 & e^{j \omega_1} & 0 & \cdots & 0 \\
        \vdots & \vdots & & \ddots   & \vdots \\
        0 & 0 & 0 &\cdots  &e^{j \omega_{2M-1}}
    \end{bmatrix}^k
     \begin{bmatrix}
        1 & e^{j \omega_0} & \cdots & e^{j \omega_0 (l-1)}\\
        1 & e^{j \omega_1} & \cdots & e^{j \omega_1 (l-1)}\\
        \vdots & \vdots & \ddots & \vdots \\
        1 & e^{j \omega_{2M-1}} & \cdots & e^{j \omega_{2M-1} (l-1)}
    \end{bmatrix}, \\
    \mbf{Y} = & 
    \begin{bmatrix}
        e^{j \omega_0} & 0 & 0 & \cdots & 0 \\
        0 & e^{j \omega_1} & 0 & \cdots & 0 \\
        \vdots & \vdots & & \ddots   & \vdots \\
        0 & 0 & 0 &\cdots  &e^{j \omega_{2M-1}}
    \end{bmatrix}^k
    \begin{bmatrix}
        e^{j \omega_0 l} \\
        e^{j \omega_1 l} \\
        \vdots \\
        e^{j \omega_{2M-1} l} \\
    \end{bmatrix}.
\end{align*}
Therefore the parameters $\bar{\boldsymbol{\alpha}}$ are defined such that,
\begin{align}
    \bar{\boldsymbol{\alpha}} &:= \mbf{V}^{\dagger} \mbf{Y}, \nonumber\\
                  &=  \begin{bmatrix}
        1 & e^{j \omega_0} & \cdots & e^{j \omega_0 (l-1)}\\
        1 & e^{j \omega_1} & \cdots & e^{j \omega_1 (l-1)}\\
        \vdots & \vdots & \ddots & \vdots \\
        1 & e^{j \omega_{2M-1}} & \cdots & e^{j \omega_{2M-1} (l-1)}
    \end{bmatrix}^{\dagger}
    \begin{bmatrix}
        e^{j \omega_0 l} \\
        e^{j \omega_1 l} \\
        \vdots \\
        e^{j \omega_{2M-1} l} \\
    \end{bmatrix} \label{Eq:AR_model_parameters},
\end{align}
where, $\bar{\boldsymbol{\alpha}} = [\alpha_0, \alpha_1, \dots, \alpha_{l-1}]^{\mathsf{T}}$. Here $\bar{\boldsymbol{\alpha}}$ is independent of the time $k$, and hence we obtain a time-invariant autoregressive model for the trajectory of the system: 
\begin{align}
    \mbf{x}(k+l) & = \alpha_0 \mbf{x}(k) + \alpha_1 \mbf{x}(k+1) + \cdots +\alpha_{l-1} \mbf{x}(k+l-1) \label{Eq:AR_model}.
\end{align}
Eq.~\eqref{Eq:AR_model} shows that the system response can be written as a linear autoregressive (AR) model using a finite history of states from the past. And based on the rank of the matrix $\mbf{V}$, the length of the history $l$ or the number of time delays should be such that $l\geq 2M$, where $M$ is the number of modes/frequencies in the signal. From Eq.~\eqref{Eq:AR_model}, consider $\mbf{L} \in \mathbb{R}^{n \times l}$ such that,
\begin{equation}
\begin{aligned}
    \mbf{L} &= 
    \begin{bmatrix}
         \mbf{x}(k) & \mbf{A}\mbf{x}(k) & \mbf{A}^2\mbf{x}(k) & \cdots & \cdots & \cdots & \mbf{A}^{l-1}\mbf{x}(k) 
    \end{bmatrix},  \\
    &= 
    \begin{bmatrix}
         \mbf{x}(k) & \mbf{x}(k+1) & \mbf{x}(k+2) & \cdots & \cdots & \cdots & \mbf{x}(k+l-1)
    \end{bmatrix}.
\end{aligned}
\end{equation}
Then, it is seen that,
\begin{align}
    \mbf{AL} = \mbf{LF}, \label{Eq:comp_eigen}
\end{align}
where $\mbf{F} \in \mathbb{R}^{l \times l}$ is a companion matrix,
\begin{align}
    \mbf{F} &=
    \begin{bmatrix}
        0 &  0 & \cdots & 0 &\alpha_0 \\
        1 &  0 & \cdots & 0 &\alpha_1 \\
        \vdots & \vdots & \ddots & \vdots & \vdots\\
        0 & \cdots & \cdots & 1 & \alpha_{l-1}
    \end{bmatrix}.
\end{align}
$\mbf{F}$ is related to the Vandermonde matrix $\mbf{V}$ in the following manner, $\mbf{F} = \mbf{V^{\dagger} D V}$, where $\mbf{D} = diag(e^{j \omega_0}, e^{j \omega_1}, \cdots, e^{j \omega_{2M-1}})$ and $\dagger$ is the Moore-Penrose inverse.
The eigenvalues of $\mbf{F}$ are the subset of the eigenvalues of $\mbf{A}$ if,
\begin{align}
    \mbf{Fa} = \lambda \mbf{a},
\end{align}
and using Eq.~\eqref{Eq:comp_eigen}, it can be verified $\mbf{v} = \mbf{La}$ is an eigenvector of $\mbf{A}$ with eigenvalue $\lambda$ ~\cite{rowley2009spectral}. 
In practice, the assumption that $\mbf{X}_k$ is full rank is generally not true, as the data matrices $\mbf{X}_k$ and $\mbf{X}_{k+1}$ are typically rank deficient. This is especially true when the data stems from experiments contaminated with noise, precision or rounding errors, and other uncertainties. SVD-based DMD offers a more robust approach to determining the AR parameters by operating in a lower-dimensional space spanned by the dominant modes of the dataset. Algo.~\ref{algo:DMD} details this process. This work used Hankel reconstruction or time-delayed embedding to improve the spatial resolution of the snapshot matrices and therefore allow the modeling of trajectories with a large number of frequencies via an autoregressive approach.

Hankel reconstruction involves recursively arranging state observations to enhance the spatial dimensions of a dataset. This technique has demonstrated its ability to represent ergodic attractors in nonlinear dynamical systems~\cite{takens1981detecting}. It is widely applied across various fields, including signal analysis and forecasting~\cite{box2015time,brunton2017chaos,arbabi2017ergodic,le2017higher}. Considering the state matrix in Eq.~\eqref{Eq:state_matrix}, the Hankel matrix $\mbf{H}_k$ can be constructed as,
\begin{align}
\mbf{H}_k &= 
    \left[\begin{array}{c}
\mathbf{X}_{k} \\
\mathbf{X}_{k+1} \\
\vdots \\
\mathbf{X}_{k+l-2} \\
\mathbf{X}_{k+l-1}
\end{array}\right] \in \mathbb{R}^{nl \times s}, \label{Eq:data_matrix}
\end{align}
where $l$ is the delay embedding dimension and,
\begin{equation}
\mathbf{X}_{k+l} \approx \mathbf{A}_0 \mathbf{X}_{k}+\mathbf{A}_1 \mathbf{X}_{k+1}+\ldots+\mathbf{A}_{l-1} \mathbf{X}_{k+l-1},
\end{equation}
where note that this embedding is precisely the AR model developed to model a periodic orbit of a nonlinear system.
The resulting first-order problem is ~\cite{nayak2023fly},
\begin{equation}
    \mbf{H}_{k+1} = \mbf{F}_{H_k}^{H_{k+1}} \mbf{H}_{k}, \ \text{or},
\end{equation}
\begin{equation}
\left[\begin{array}{c}
\mathbf{X}_{k+1} \\
\mathbf{X}_{k+2} \\
\vdots \\
\mathbf{X}_{k+l-1} \\
\mathbf{X}_{k+l}
\end{array}\right] \approx\left[\begin{array}{ccccc}
\mathbf{0}_n & \mathbf{I}_n & \ldots & \mathbf{0}_n & \mathbf{0}_n \\
\mathbf{0}_n & \mathbf{0}_n & \ldots & \mathbf{0}_n & \mathbf{0}_n \\
\vdots & \vdots & \ddots & \vdots & \vdots \\
\mathbf{0}_n & \mathbf{0}_n & \ldots & \mathbf{0}_n & \mathbf{I}_n \\
\mathbf{A}_0 & \mathbf{A}_{1} & \ldots & \mathbf{A}_{l-2} & \mathbf{A}_{l-1}
\end{array}\right]\left[\begin{array}{c}
\mathbf{X}_{k} \\
\mathbf{X}_{k+1} \\
\vdots \\
\mathbf{X}_{k+l-2} \\
\mathbf{X}_{k+l-1}
\end{array}\right],
\end{equation}
where $\mbf{0}_n$ and $\mbf{I}_n$ are zero and identity matrices of dimension $n \times n$. And $\mbf{F}_{H_k}^{H_{k+1}}$ is the block matrix that relates the two Hankel matrices $\mbf{H}_k$ and $\mbf{H}_{k+1}$. For cases when $\mbf{X}_{k+1} = \mbf{AX}_k$ exactly, then $\mbf{A}_0 = \mbf{A}, \mbf{A}_1 = \mbf{A}^2, \cdots, \mbf{A}_{l-1} = \mbf{A}^l$. Both Hankel-DMD and regular DMD produce the same set of eigenvalues and modes after truncation \cite{nayak2023fly}.
To use Hankel-DMD the only change is that $\mbf{H}_k$ and $\mbf{H}_{k+1}$ are used as inputs to the DMD algorithm. First, the value of $l$ is guessed to build the data matrix $\mbf{H}_k$ such that the number of columns is greater than the number of rows, i.e. $s > n l$. Then the minimum number of time delays is given by the rank of the data matrix $\mbf{H}_k$. If $\mbf{H}_k$ has full row rank, then $l$ is iteratively increased until the matrix becomes row rank deficient. The following section presents results from the experimental analysis of the three candidate periodic orbits.

\begin{algorithm}[h]
\caption{DMD Algorithm}
\label{algo:DMD}
\textcolor{blue}{(Goal)} To identify an optimal linear operator $\mathbf{A}$ such that:
$\mathbf{X}_{k+1} \approx \mathbf{A}\mathbf{X}_k$ or $\mathbf{X}_{k+l} \approx \mathbf{A}^l\mathbf{X}_k$  
and predict the evolution of the trajectory at a future time instant $k$.
\begin{enumerate}
    \item Form snapshot matrices $\mbf{X}_k$ and $\mbf{X}_{k+1}$, where $\mbf{X}_{k+1}$ is one-time snapshot ahead of the original snapshot matrix $\mbf{X}_k$
    \begin{align}
    \mbf{X}_k &= 
    \begin{bmatrix}
        \mbf{x}(k) & \mbf{x}(k+1) & \mbf{x}(k+2) & \cdots &  \mbf{x}(k+s-1)\\
    \end{bmatrix},\label{Eq:state_matrix}\\
    \mbf{X}_{k+1} &= 
    \begin{bmatrix}\label{eq:dmd_snap_mat}
        \mbf{x}(k+1) & \mbf{x}(k+2) & \mbf{x}(k+3) & \cdots &  \mbf{x}(k+s)\\
    \end{bmatrix}.
    \end{align}
    It is assumed that $s \gg l$, i.e., the training window is much larger than the time delay dimension required. This additional temporal length improves the spatial resolution via the Hankel-DMD method~\cite{narayanan2024predictive}.

    \item Compute the SVD of snapshot matrix $\mathbf{X}_k$:
    \begin{align}
        \mathbf{X}_k &= \mathbf{U} \mathbf{\Sigma} \mathbf{V}^{H} \approx \mathbf{\Tilde{U}} \mathbf{\Tilde{\Sigma}} \mathbf{\Tilde{V}}^{H},
    \end{align}
    where $\mathbf{U}$ contains the left singular vectors, $\mathbf{\Sigma}$ is the diagonal matrix of singular values, and $\mathbf{V}$ contains the right singular vectors.

    \item Calculate the reduced order space projection:
    \begin{align}
        \mathbf{A} &= \mathbf{X}_{k+1} \mathbf{X}_k^{\dagger} = \mathbf{X}_{k+1} \mathbf{\Tilde{V}} \mathbf{\Tilde{\Sigma}}^{-1} \mathbf{\Tilde{U}}^{H}, \\
        \mathbf{\Tilde{A}} &= \mathbf{\Tilde{U}}^{H} \mathbf{A} \mathbf{\Tilde{U}} = \mathbf{\Tilde{U}}^{H} \mathbf{X}_{k+1} \mathbf{\Tilde{V}} \mathbf{\Tilde{\Sigma}}^{-1}.
    \end{align}

    \item Perform the eigendecomposition of $\mathbf{\Tilde{A}}$:
    \begin{align}
        \mathbf{\Tilde{A}} \mathbf{W} &= \mathbf{W} \mathbf{\Lambda},
    \end{align}
    where $\mathbf{W}$ contains the eigenvectors and $\mathbf{\Lambda}$ is the diagonal matrix of eigenvalues.

    \item Compute the DMD modes:
    \begin{align}
        \mathbf{Z} &= \mathbf{X}_{k+1} \mathbf{\Tilde{V}} \mathbf{\Tilde{\Sigma}}^{-1} \mathbf{W}. ^{*}
    \end{align}

    \item Obtain the DMD predicted state at time $k$:
    \begin{align}
        \hat{\mathbf{x}}(k) &= \mathbf{Z} \mathbf{\Lambda}^{\frac{k}{\Delta k}} \mathbf{b},
    \end{align}
    where $\mathbf{b} = \mathbf{Z}^{\dagger} \mathbf{x}(0)$ is the scaling factor of DMD modes. \\
    
    *Note: This formulation follows \textit{exact}-DMD and differs from the original formulation which defines $\mbf{Z} = \bsym{\Tilde{U}W}$ as the projected DMD modes.
\end{enumerate}
\end{algorithm}

\section{Results}
\label{Sec:Res}
The three orbits considered in this work include an L1-Halo orbit, an L1-L2 butterfly orbit, and a resonant (2:1) orbit. All orbits were generated within the \ac{cr3bp} framework. The system was non-dimensionalized using the following physical properties: mass ratio $\frac{M_E}{M_E + M_M} = 1.2150e-2$, length unit $\text{(LU)} = 389703 \ $km, time unit $\text{(TU)} = 382981 \ $sec and radius of the moon $R_M = 1737.1 \ km$. Here, LU is the distance between the two primaries, Earth and Moon, and TU is the inverse of the relative angular frequency between the primaries.
The normalized 2-norm error ($\bsym{\epsilon}$) is used as the error metric to assess DMD training and prediction accuracy. It is calculated between the true state data $\mbf{x}(k)$ and the estimates $\bsym{\hat{x}}(k)$ as given by, 
\begin{equation}
\bsym{\epsilon}({\mbf{x}}, \bsym{\hat{x}})=\frac{\|\bsym{\hat{x}}(k)-{\mbf{x}}(k)\|_2}{\max(\mbf{x})}.
\end{equation}
The experimental setup proceeded as follows: A constant training window size of $W_{\text{TRN}} = 10$ periods and a sampling interval ($\Delta k$) following the Nyquist rate was initially chosen. During this phase, the prediction window size was set to $W_{\text{PRED}}=0$ to prevent extrapolation by the DMD method. An exhaustive search was conducted to determine the optimal number of time delays, varying them across multiple runs of the DMD algorithm and selecting the configuration with the minimum error $\bsym{\epsilon}$. Similarly, the training window size was adjusted while keeping the number of time delays constant. Finally, the prediction window size was determined based on the established values of the previous parameters.
The selection of the sampling interval $\Delta k$ is primarily experimental. This involves an ad-hoc method that observes how changes in the sampling interval affect both the error metric and the frequency content of the reconstruction. Using a Fourier basis, as discussed in Sec.~\ref{Sec:Th}, requires at least one period of training data to establish the corresponding linear AR model, as depicted in Eq.~\eqref{Eq:Fourier_series}. Our approach enforces this requirement and employs an experimental method to determine the optimal training window size that minimizes the normalized error $\bsym{\epsilon}$. The fundamental frequency and period of the orbit are obtained via the eigenvalues of the surrogate system matrix  $\mbf{\Tilde{A}}$ found by DMD. 
The stability of the orbit is determined by plotting the eigenvalues of the system matrix $\mbf{\Tilde{A}}$ and confirming that they lie within the unit circle.
Further, the period is verified by counting the time between successive crossings of the orbit with the coordinate hyperplanes.

\subsection{L1 - Halo Orbit}

\begin{table}[htbp]
    \fontsize{10}{10}\selectfont
    \caption{L1 - Halo orbit initial conditions in the rotating coordinate frame (nondimensionalized)}
    \label{tab:L1_init}
    \centering 
    \begin{tabular}{c|c|c|c|c|c} 
        \hline 
        $\mbf{x}$ & $\mbf{y}$ & $\mbf{z}$ & $\dot{\mbf{x}}$ & $\dot{\mbf{y}}$ & $\dot{{z}}$ \\
        \hline 
        8.7592e-1 & -1.5903e-26 & 1.9175e-1 & -2.9302e-14 & 2.3080e-1 & 7.36497e-14 \\
        \hline
    \end{tabular}
\end{table}

\begin{figure}[h]
\begin{minipage}{.5\linewidth}
\centering
\subfigure[]{\label{fig:L1HALO_main:a}\includegraphics[scale=0.27]{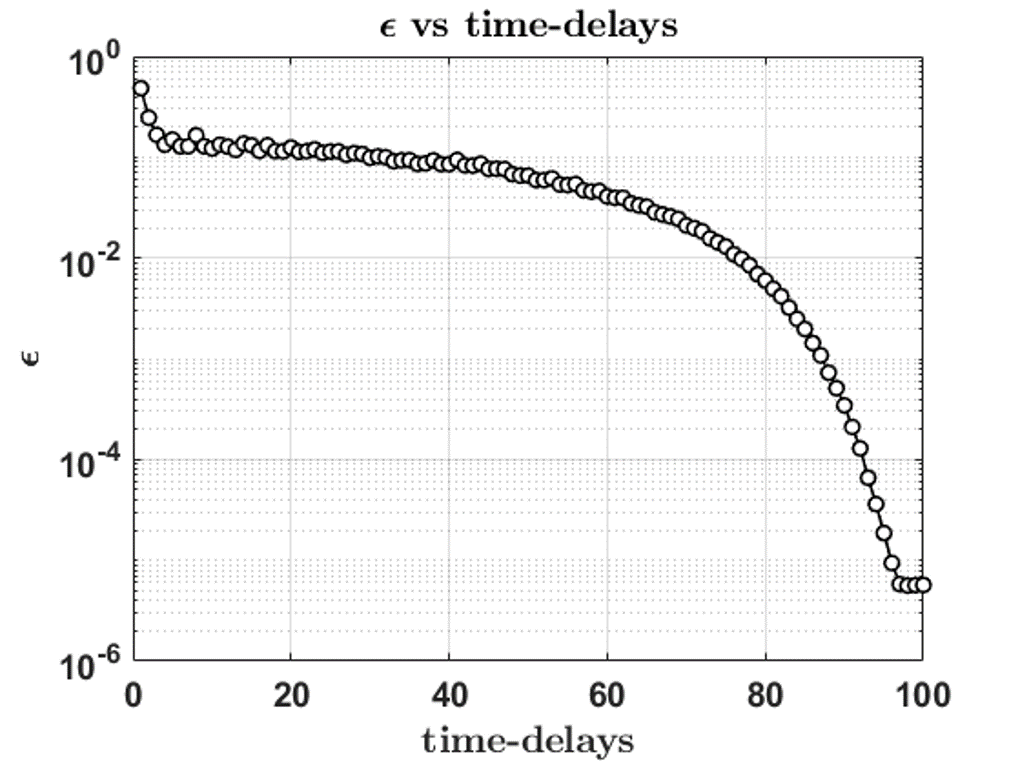}}
\end{minipage}%
\begin{minipage}{.5\linewidth}
\centering
\subfigure[]{\label{fig:L1HALO_main:b}\includegraphics[scale=0.27]{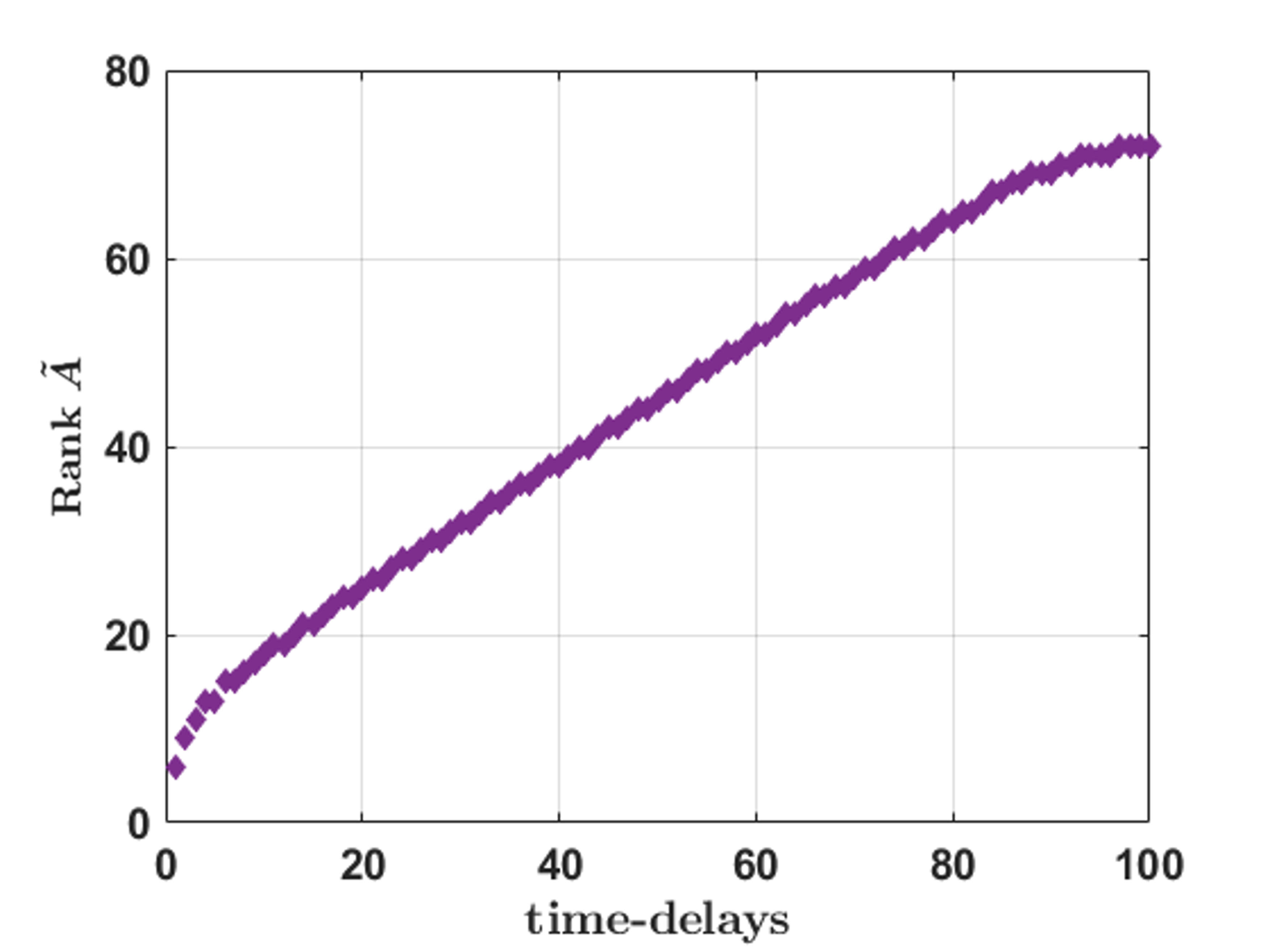}}
\end{minipage}
\begin{minipage}{.5\linewidth}
\centering
\subfigure[]{\label{fig:L1HALO_main:c}\includegraphics[scale=0.27]{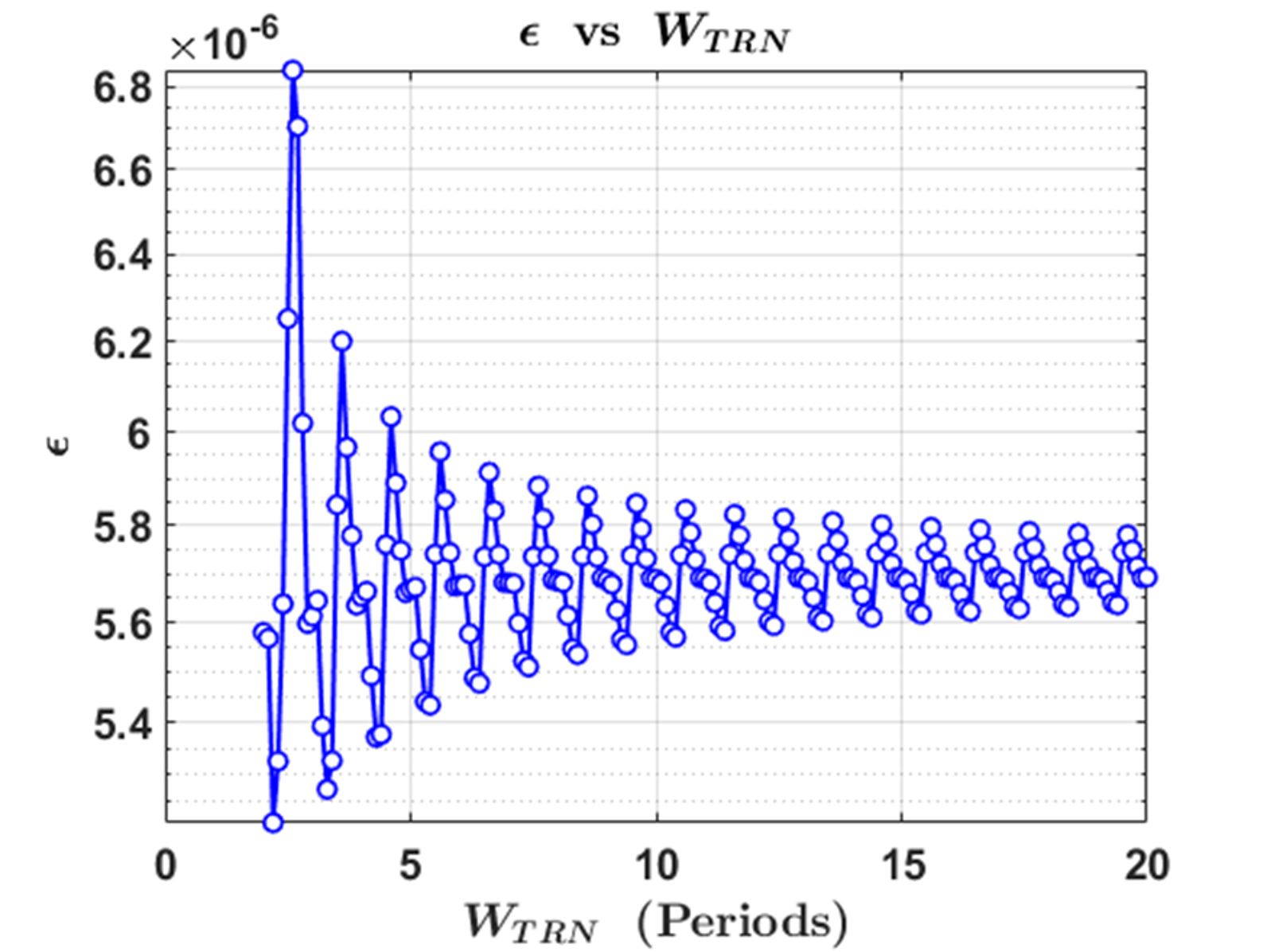}}
\end{minipage}
\begin{minipage}{.5\linewidth}
\centering
\subfigure[]{\label{fig:L1HALO_main:d}\includegraphics[scale=0.27]{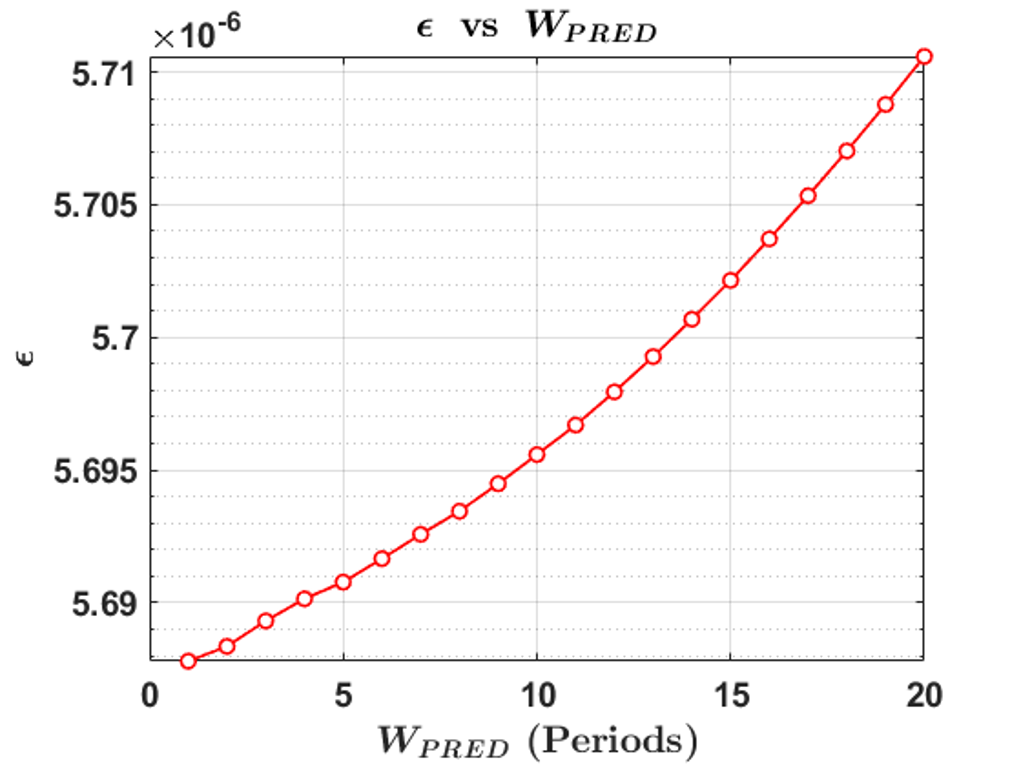}}
\end{minipage}
\caption{L1 - Halo orbit}
\label{fig:L1HALO}
\end{figure}

This scenario utilized a sampling interval of $\Delta k = 21 \ TU$. As shown in Fig.~\eqref{fig:L1HALO_main:a}, a minimum of 97 time delays is necessary to achieve the smallest reconstruction error at this sampling rate. 
Here, it is observed that increasing the number of time delays raises the rank of the $\Tilde{\mbf{A}}$ matrix. The error value $\bsym{\epsilon}$ stabilizes once the matrix achieves full rank, as discussed in Sec.~\ref{Sec:Th}.
Figure~\eqref{fig:L1HALO_main:b} indicates that with 97 time delays or more, the reduced-order system matrix has a rank of 72, highlighting the presence of 36 dominant frequencies that must all be captured for accurate modeling and prediction. Figures.~\eqref{fig:L1HALO_main:c} and~\eqref{fig:L1HALO_main:d} show error over increasing training window lengths and prediction error over $20$ periods for $W_{\text{TRN}} = 10$ periods. The error is not highly sensitive to the training window size. As expected, the prediction accuracy diminishes further into the future we estimate. Here, it is observed that the error $\bsym{\epsilon}$ is not highly sensitive to the training window size but increases significantly with longer prediction horizons.

\subsection{L1/L2 - Butterfly Orbit}

\begin{table}[htbp]
    \fontsize{10}{10}\selectfont
    \caption{L1/L2 - Butterfly orbit initial conditions in the rotating coordinate frame (nondimensionalized)}
    \label{tab:L1_new_init}
    \centering 
    \begin{tabular}{c|c|c|c|c|c} 
        \hline 
        $\mbf{x}$ & $\mbf{y}$ & $\mbf{z}$ & $\dot{\mbf{x}}$ & $\dot{\mbf{y}}$ & $\dot{\mbf{z}}$ \\
        \hline 
        9.0454e-1 & -3.0043e-26 & 1.4388e-1 & -8.5657e-15 & -4.9802e-2 & -1.9332e-14 \\
        \hline
    \end{tabular}
\end{table}

\begin{figure}[h]
\begin{minipage}{.5\linewidth}
\centering
\subfigure[]{\label{fig:L12BFLY_main:a}\includegraphics[scale=0.27]{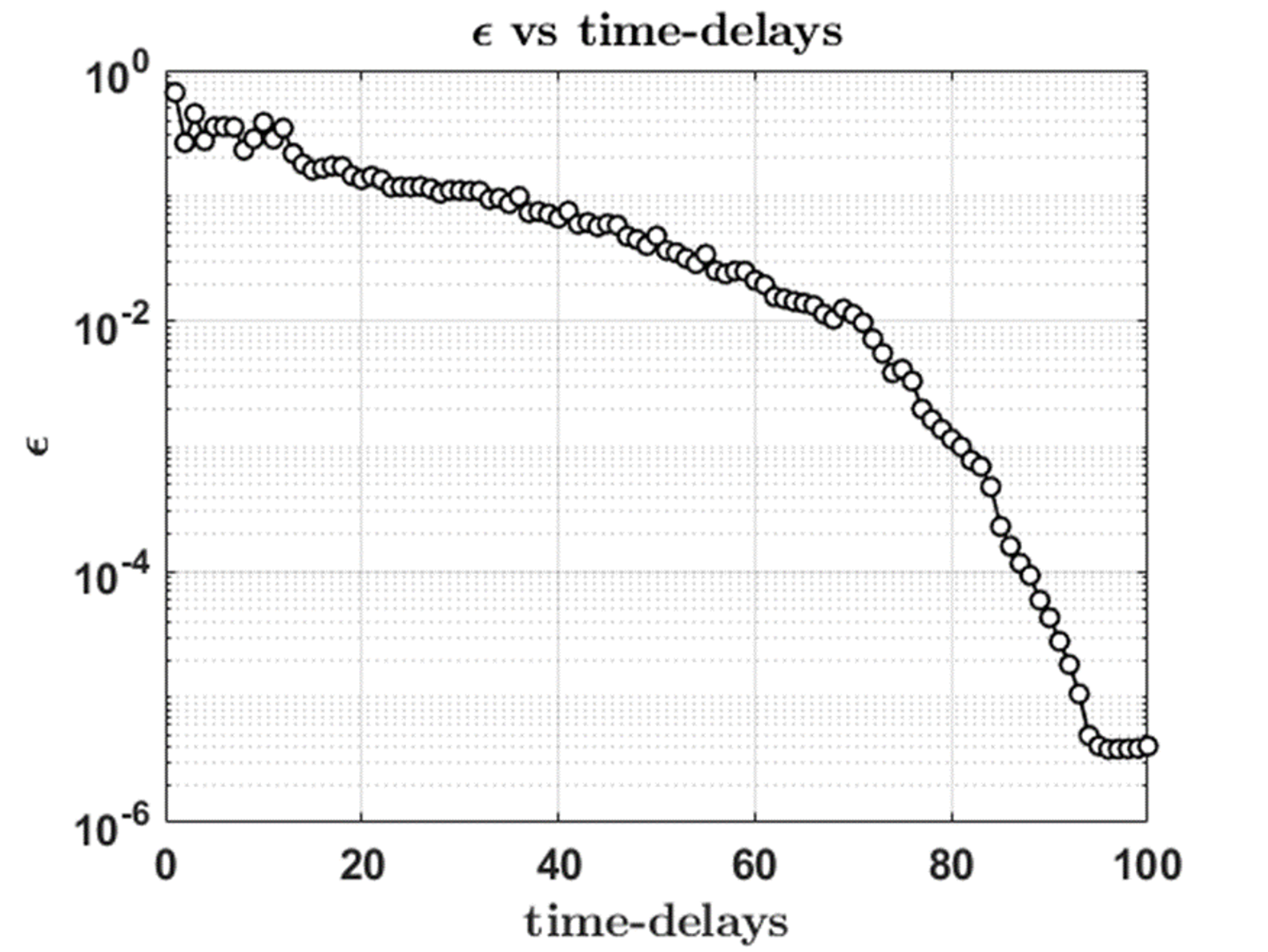}}
\end{minipage}%
\begin{minipage}{.5\linewidth}
\centering
\subfigure[]{\label{fig:L12BFLY_main:b}\includegraphics[scale=0.27]{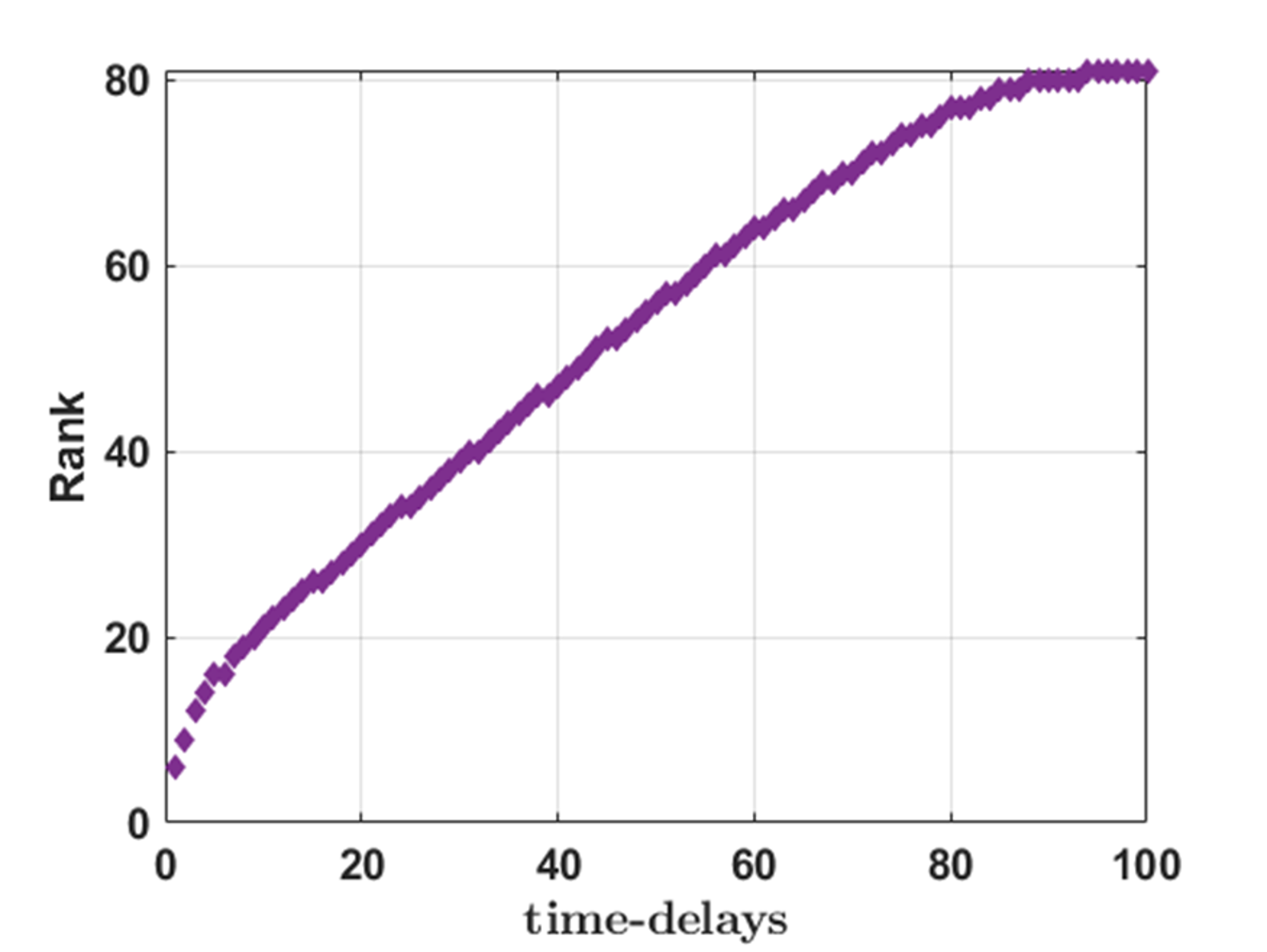}}
\end{minipage}
\begin{minipage}{.5\linewidth}
\centering
\subfigure[]{\label{fig:L12BFLY_main:c}\includegraphics[scale=0.27]{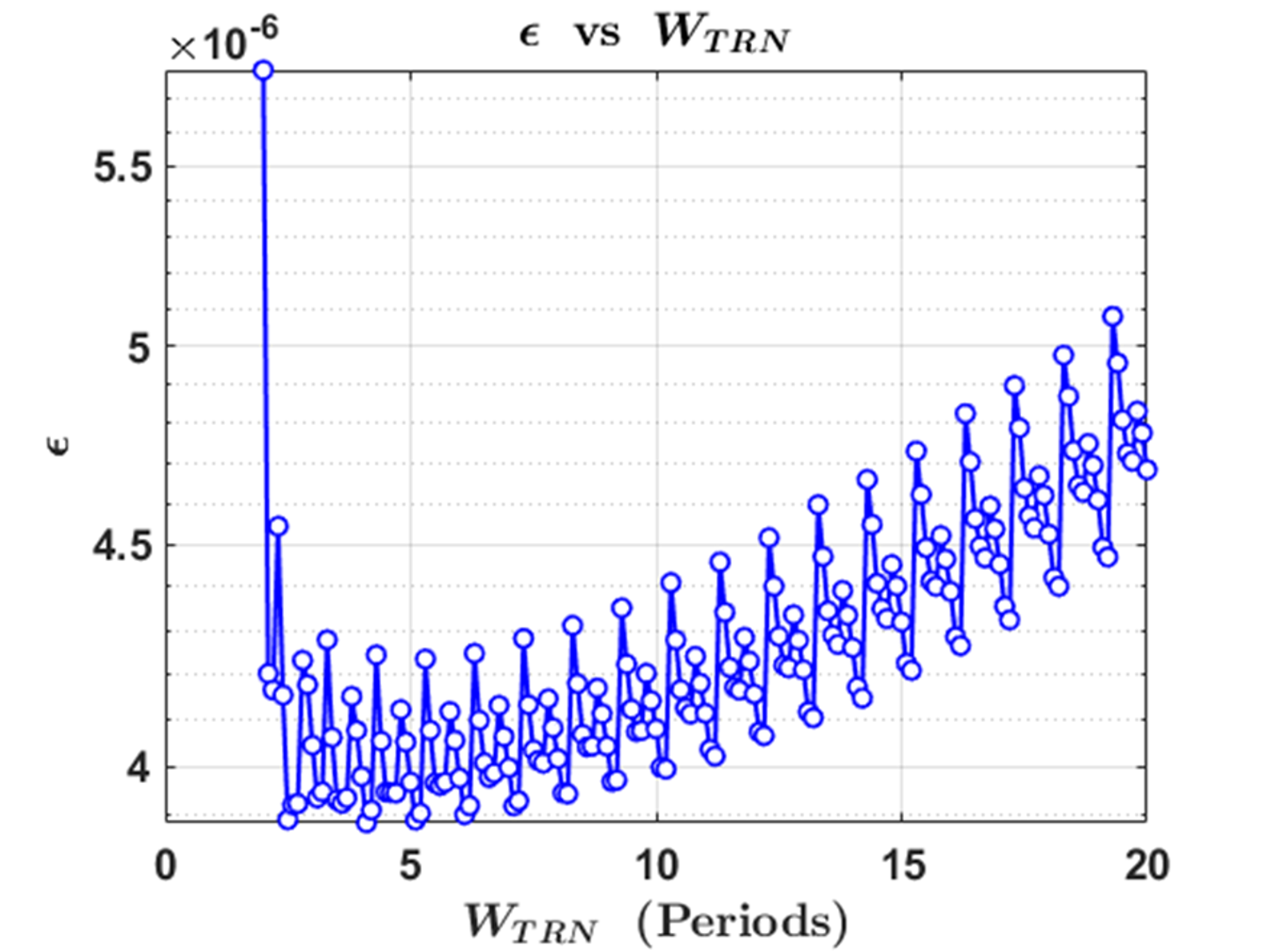}}
\end{minipage}
\begin{minipage}{.5\linewidth}
\centering
\subfigure[]{\label{fig:L12BFLY_main:d}\includegraphics[scale=0.27]{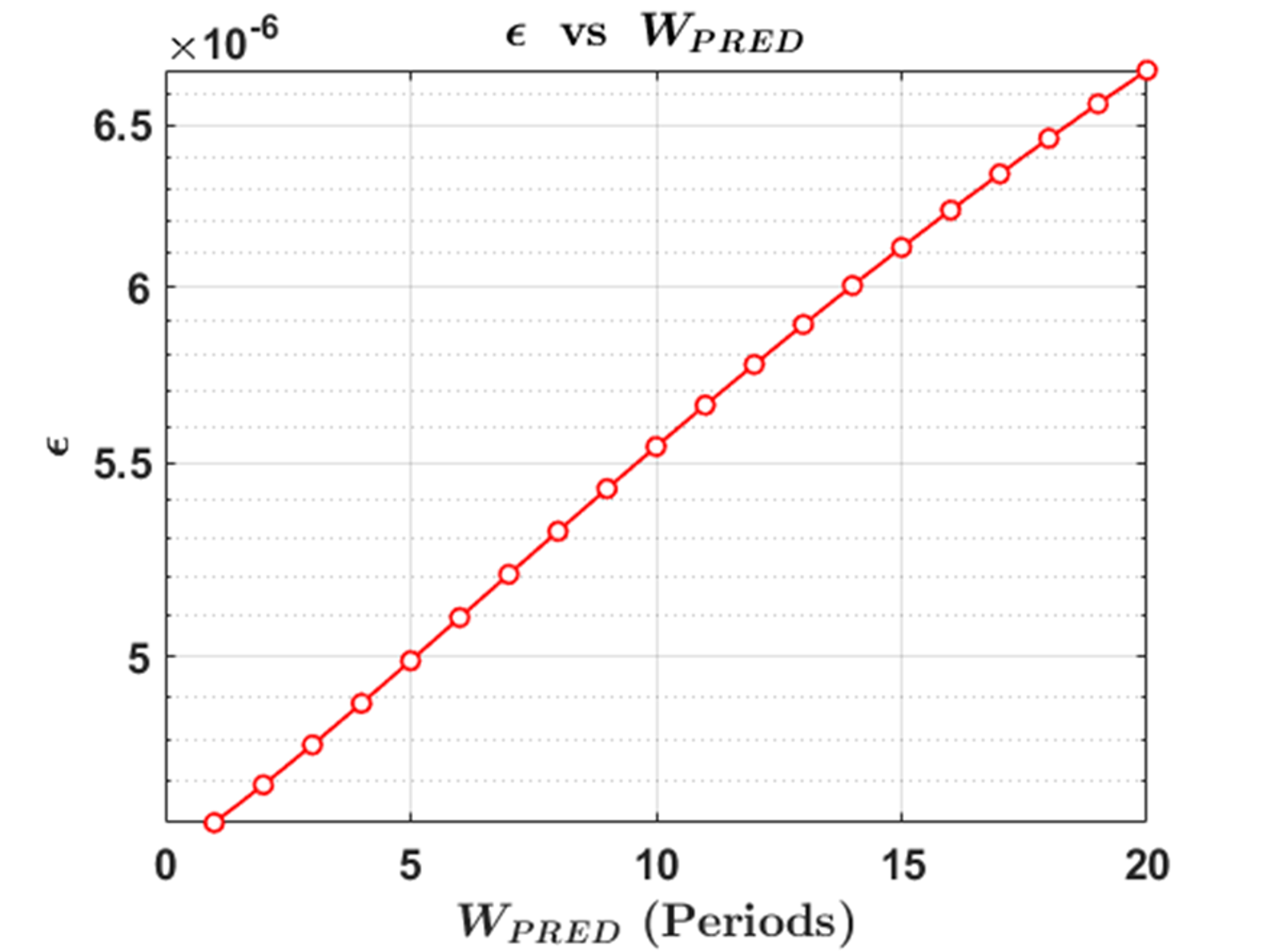}}
\end{minipage}
\caption{L1/L2 - Butterfly orbit}
\label{fig:L12BFLY}
\end{figure}

A sampling interval of $\Delta k = 18 \ TU$ was selected through iterative testing to minimize $\bsym{\epsilon}$, as discussed previously. 
Fig.~\eqref{fig:L1HALO_main:a} demonstrates that a minimum of 92 time delays is necessary to achieve the smallest reconstruction error at this sampling rate. 
In Fig.~\eqref{fig:L1HALO_main:b}, it is evident that using 92 time delays or more results in a reduced-order system matrix with a rank of 81, indicating the presence of 40 dominant frequencies and a real mode (on the real line, with no imaginary part) that must all be accurately captured for modeling and prediction.
Figs.~\eqref{fig:L1HALO_main:c} and~\eqref{fig:L1HALO_main:d} depict the trends in error with increasing training window lengths and prediction errors over 20 periods with a fixed $W_{\text{TRN}} = 10$ periods, respectively.

\subsection{Resonant (2:1) Orbit}

\begin{table}[htbp]
    \fontsize{10}{10}\selectfont
    \caption{Resonant (2:1) orbit initial conditions in the rotating coordinate frame (nondimensionalized)}
    \label{tab:L1_new_init_2}
    \centering 
    \begin{tabular}{c|c|c|c|c|c} 
        \hline 
        $\mbf{x}$ & $\mbf{y}$ & $\mbf{z}$ & $\dot{\mbf{x}}$ & $\dot{\mbf{y}}$ & $\dot{\mbf{z}}$ \\
        \hline 
        2.6656e-1 & 2.7444e-20 & -1.2692e-22 & -4.6279e-13 & 2.0683e+0 & 3.2448e-22 \\
        \hline
    \end{tabular}
\end{table}

\begin{figure}[h]
\begin{minipage}{.5\linewidth}
\centering
\subfigure[]{\label{fig:Reso_main:a}\includegraphics[scale=0.27]{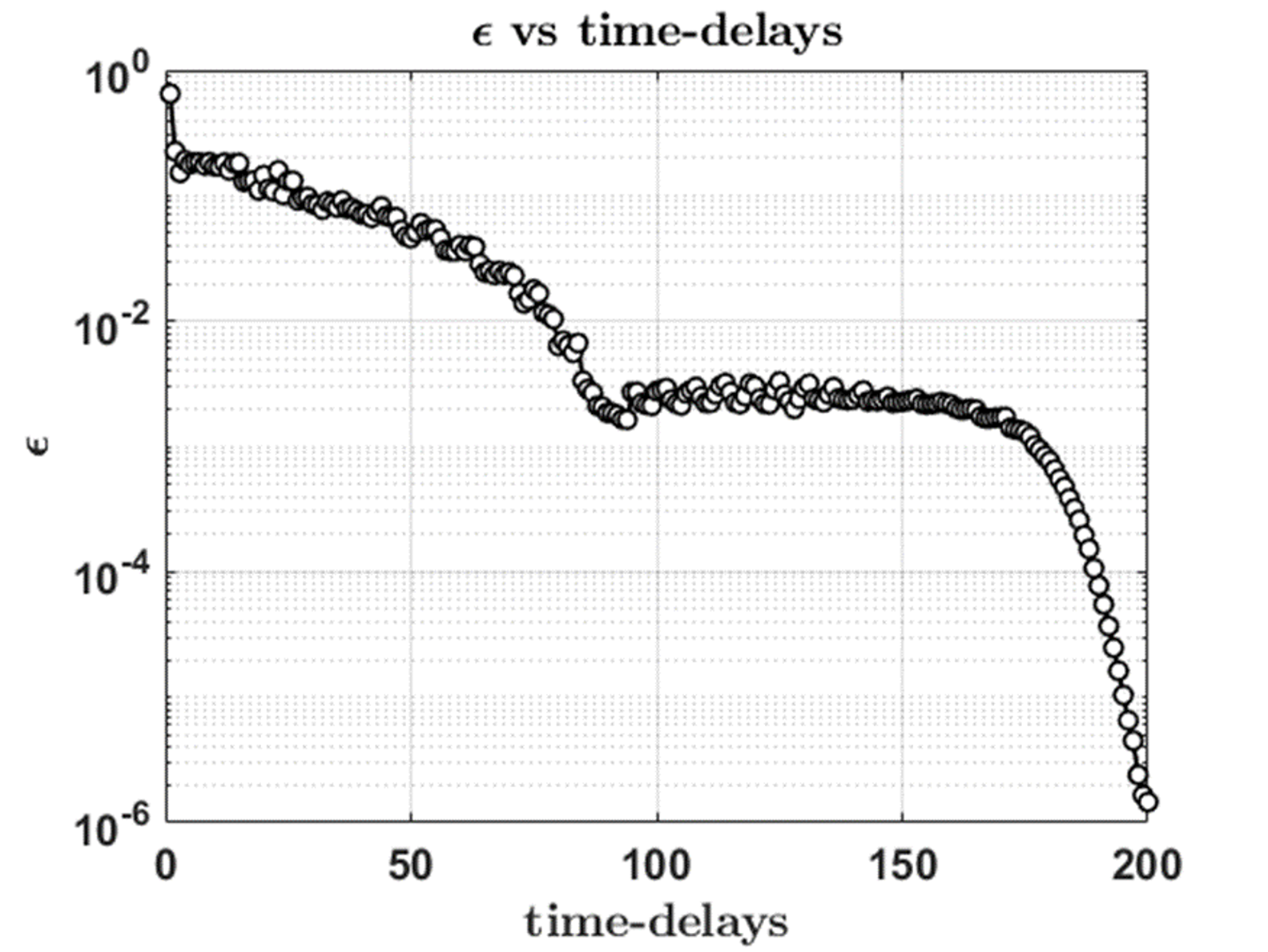}}
\end{minipage}%
\begin{minipage}{.5\linewidth}
\centering
\subfigure[]{\label{fig:Reso_main:b}\includegraphics[scale=0.27]{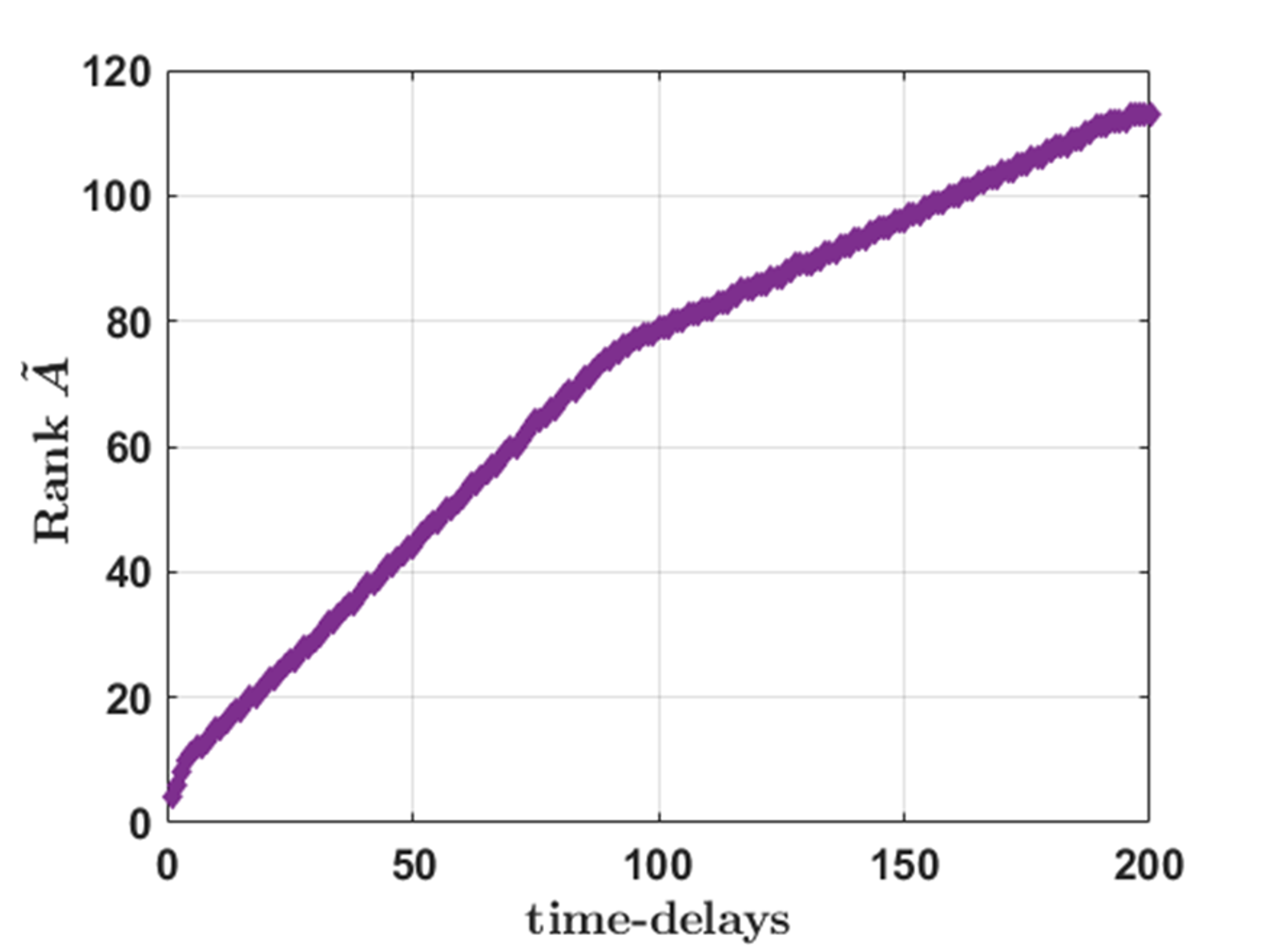}}
\end{minipage}
\begin{minipage}{.5\linewidth}
\centering
\subfigure[]{\label{fig:Reso_main:c}\includegraphics[scale=0.27]{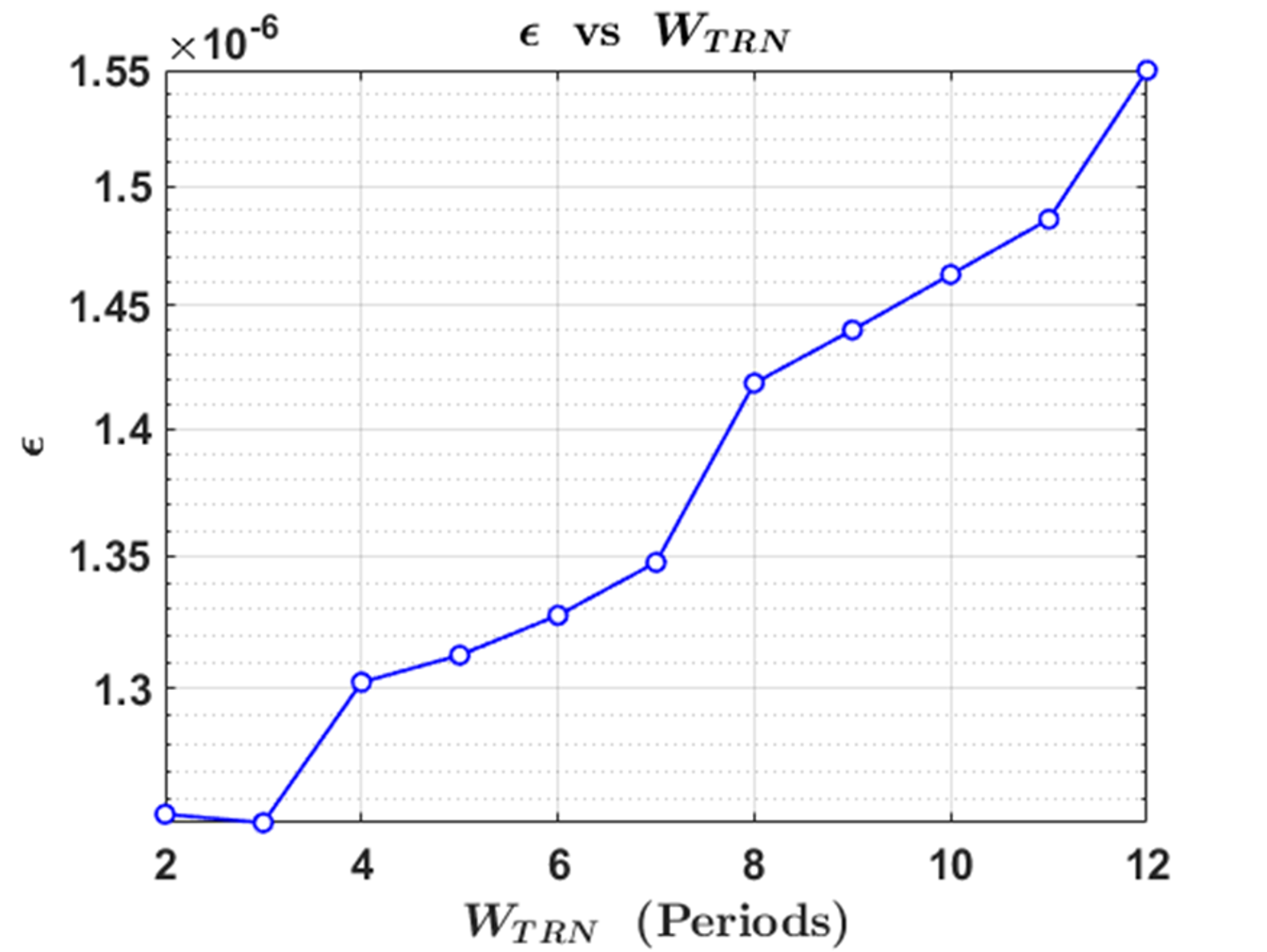}}
\end{minipage}
\begin{minipage}{.5\linewidth}
\centering
\subfigure[]{\label{fig:Reso_main:d}\includegraphics[scale=0.27]{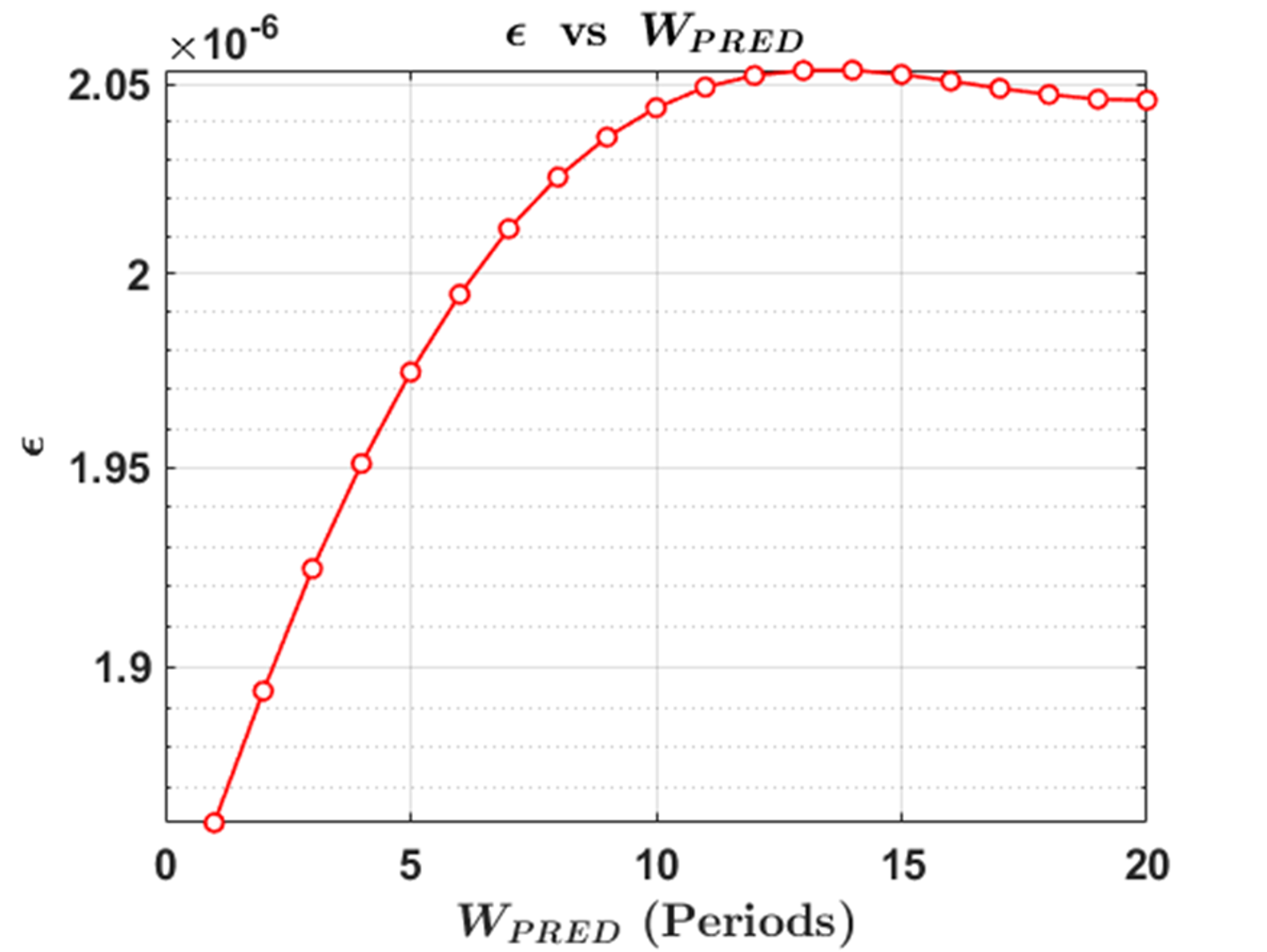}}
\end{minipage}
\caption{Resonant (2:1) orbit}
\label{fig:Reso}
\end{figure}

Similar to the previous two cases, through iterative testing to minimize $\bsym{\epsilon}$, a sampling interval of $\Delta k = 31 \ TU$ was chosen. 
Fig.~\eqref{fig:L1HALO_main:a} illustrates that at least 199 time delays are required to achieve the smallest reconstruction error at this sampling rate. 
In Fig.~\eqref{fig:L1HALO_main:b}, it is clear that using 199 time delays or more results in a reduced-order system matrix with a rank of 113, indicating the presence of 56 dominant frequencies and a real mode that must all be accurately captured for modeling and prediction.
Figures~\eqref{fig:L1HALO_main:c} and~\eqref{fig:L1HALO_main:d} show the error trends with increasing training window lengths and prediction errors over 20 periods, respectively, while maintaining a fixed $W_{\text{TRN}} = 10$ periods. The observed trends are consistent with those seen in the earlier cases.

\subsection{Spectral Comparison}

\begin{figure}[h]
\begin{minipage}{.33\linewidth}
\centering
\subfigure[]{\label{fig:Spec_main:a}\includegraphics[scale=0.18]{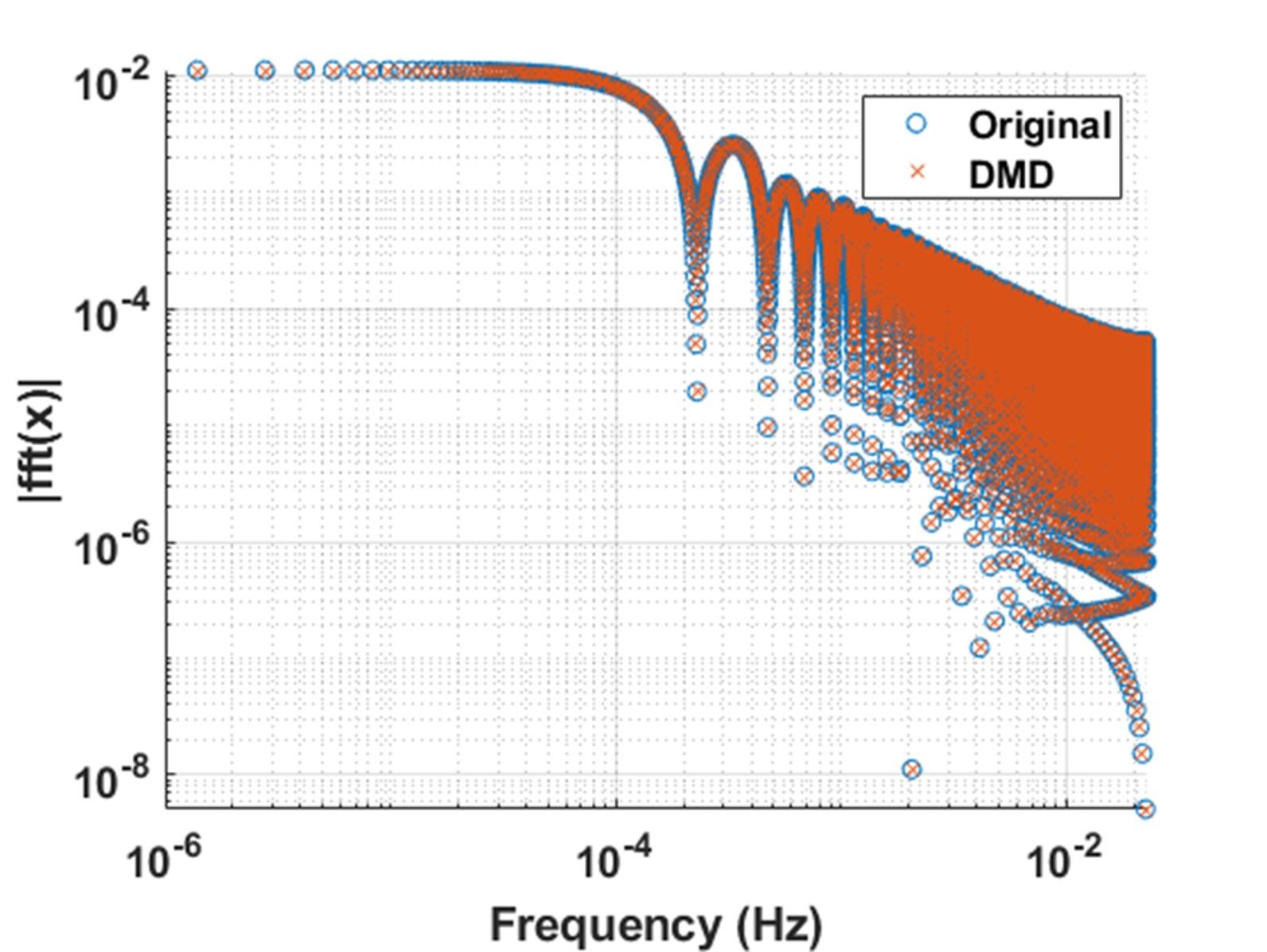}}
\end{minipage}%
\begin{minipage}{.33\linewidth}
\centering
\subfigure[]{\label{fig:Spec_main:b}\includegraphics[scale=0.18]{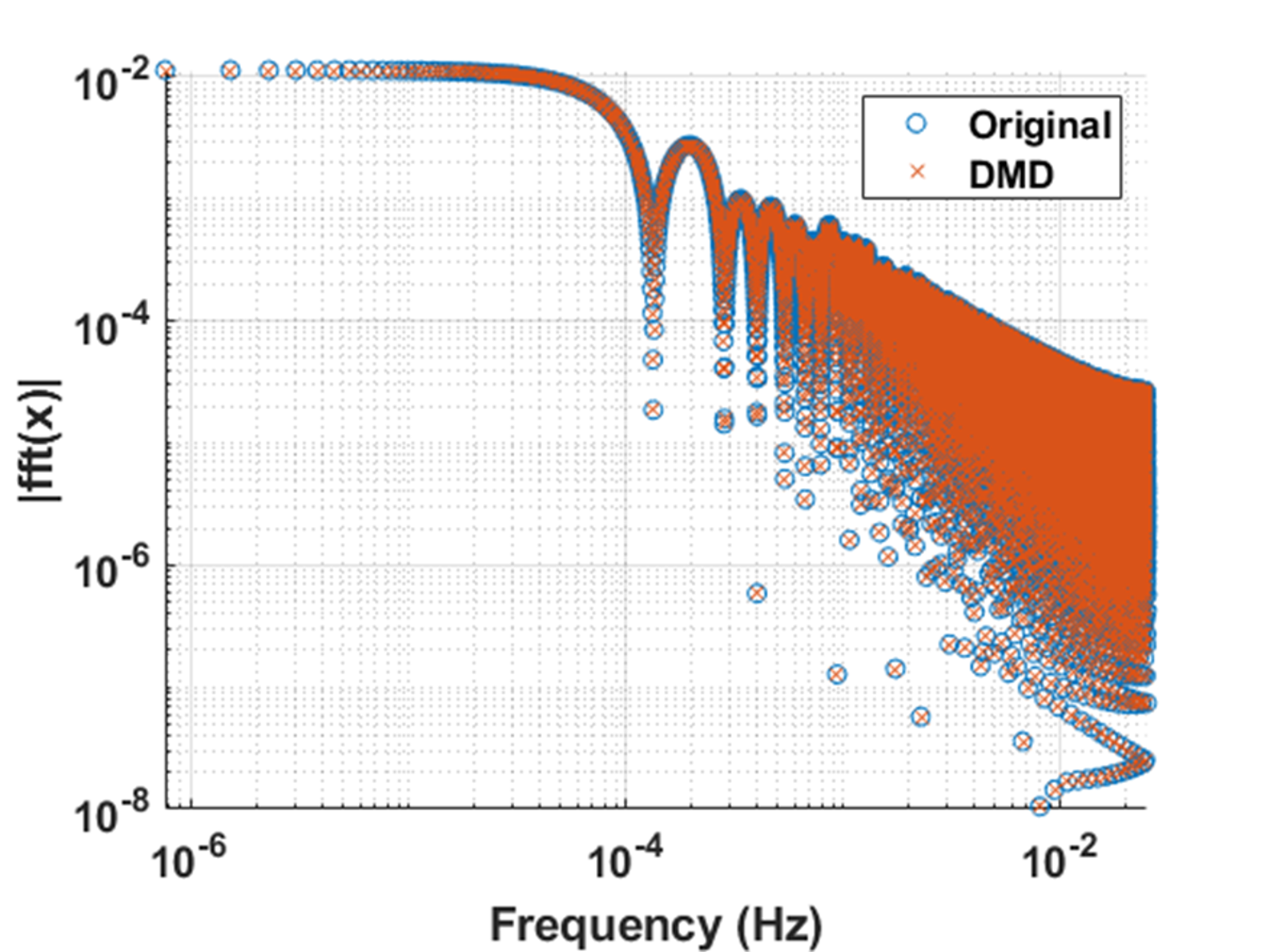}}
\end{minipage}
\begin{minipage}{.33\linewidth}
\centering
\subfigure[]{\label{fig:Spec_main:c}\includegraphics[scale=0.18]{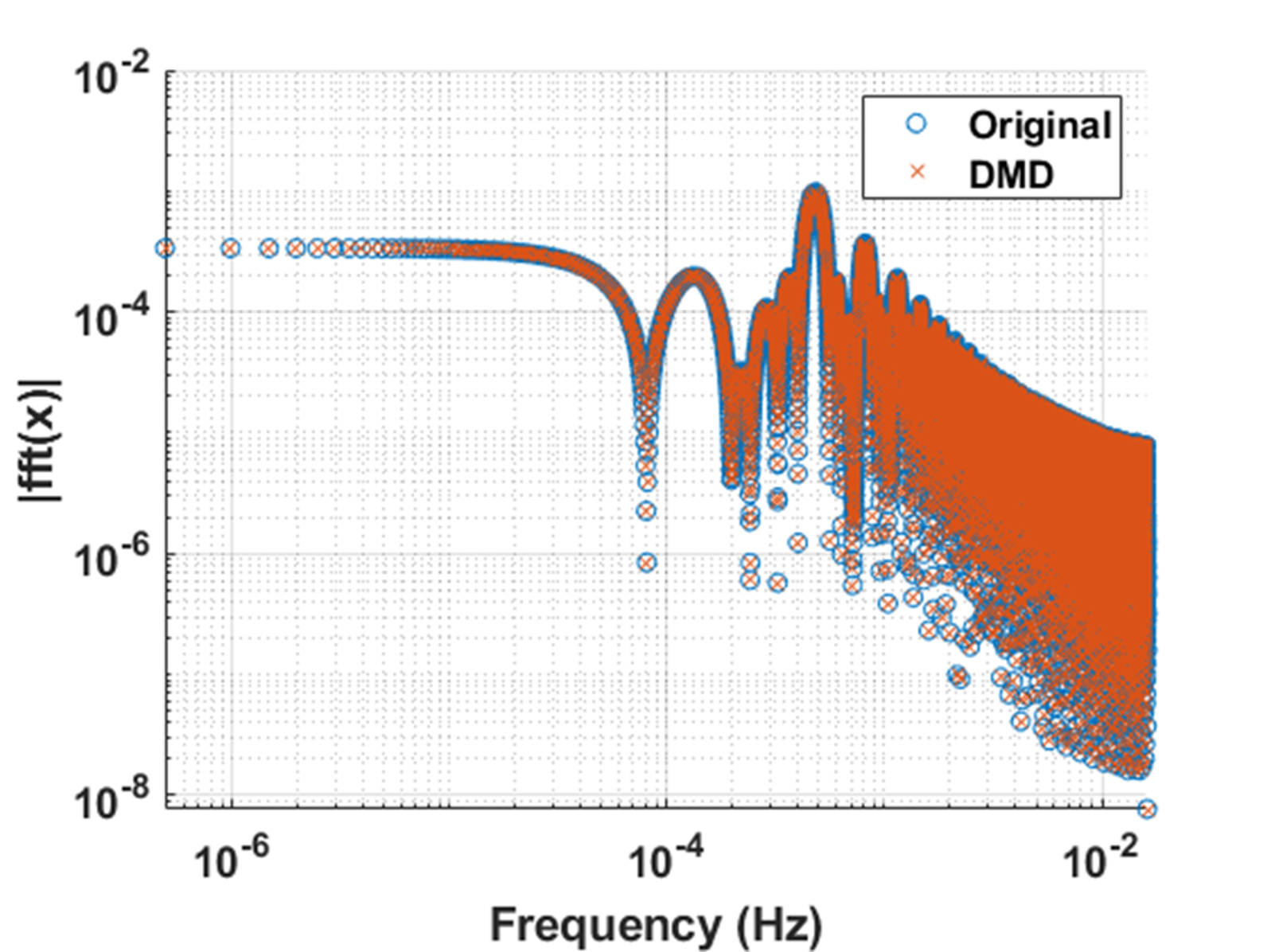}}
\end{minipage}
\caption{Spectral comparison for each orbit type: (a) L1 Halo orbit (b) L1/L2 Butterfly orbit (c) Resonant (2:1) orbit}
\label{fig:SpecComp}
\end{figure}

\begin{figure}[h]
\begin{minipage}{.33\linewidth}
\centering
\subfigure[]{\label{fig:Eig_main:a}\includegraphics[scale=0.18]{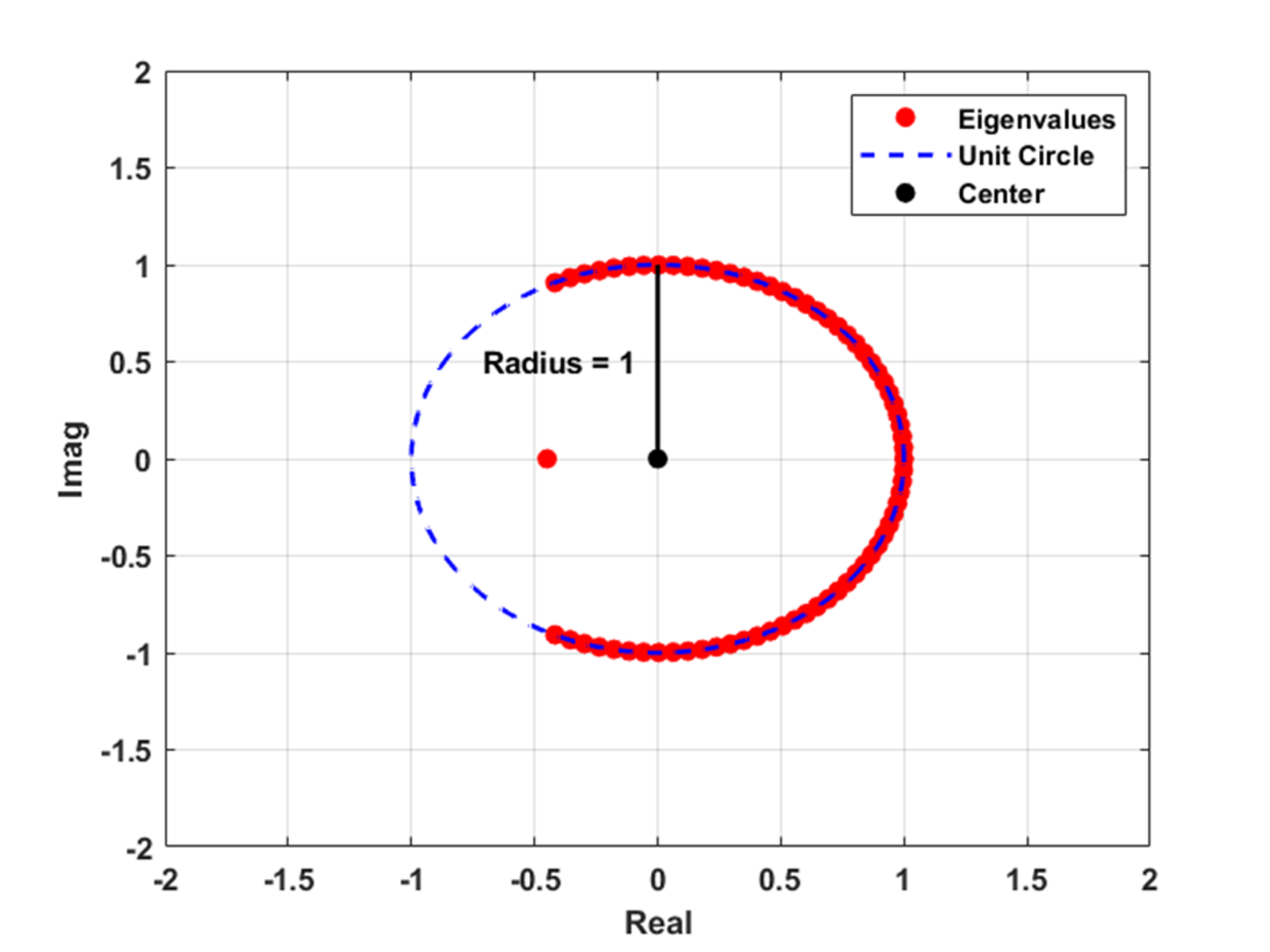}}
\end{minipage}%
\begin{minipage}{.33\linewidth}
\centering
\subfigure[]{\label{fig:Eig_main:b}\includegraphics[scale=0.18]{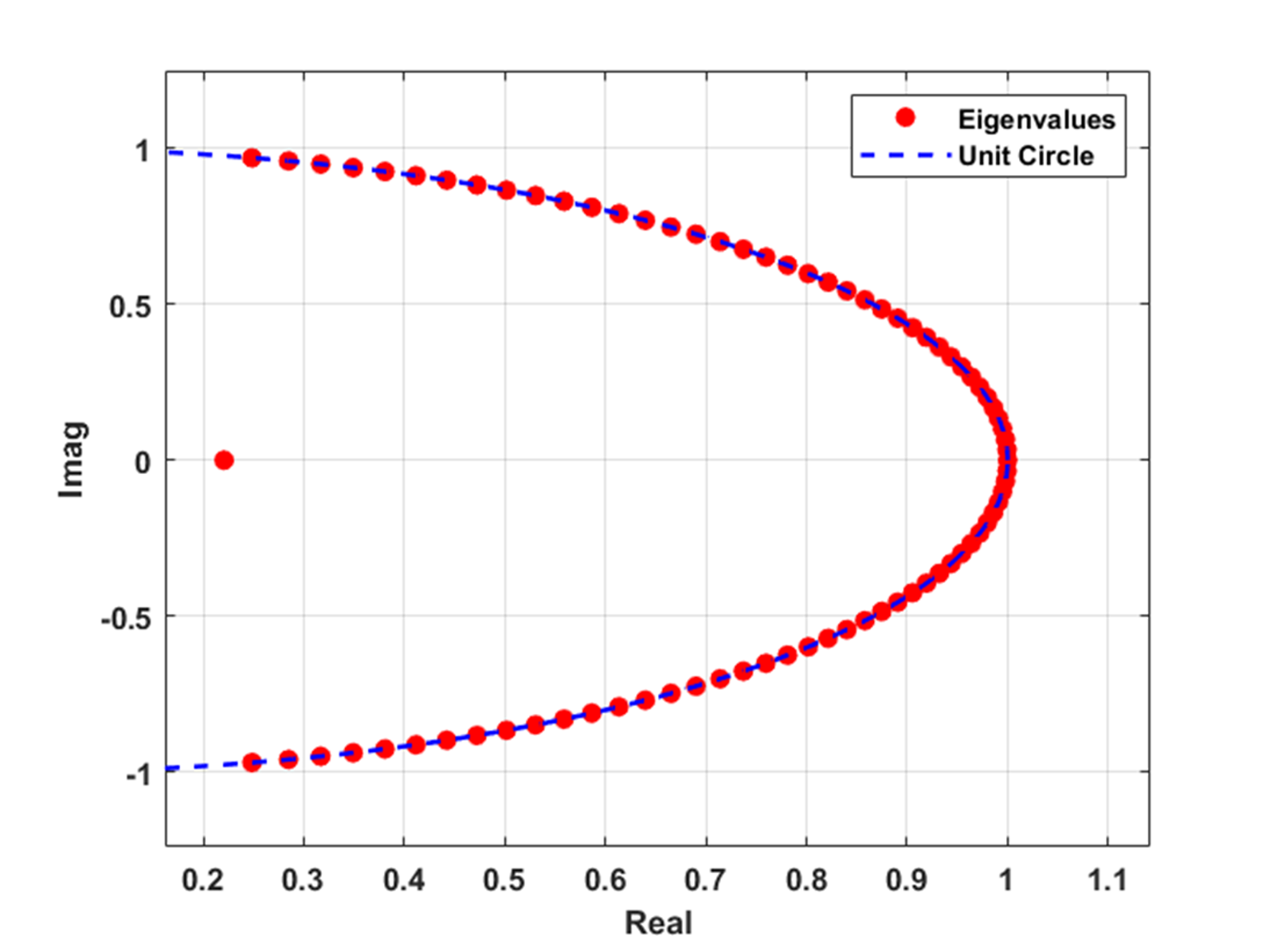}}
\end{minipage}
\begin{minipage}{.33\linewidth}
\centering
\subfigure[]{\label{fig:Eig_main:c}\includegraphics[scale=0.18]{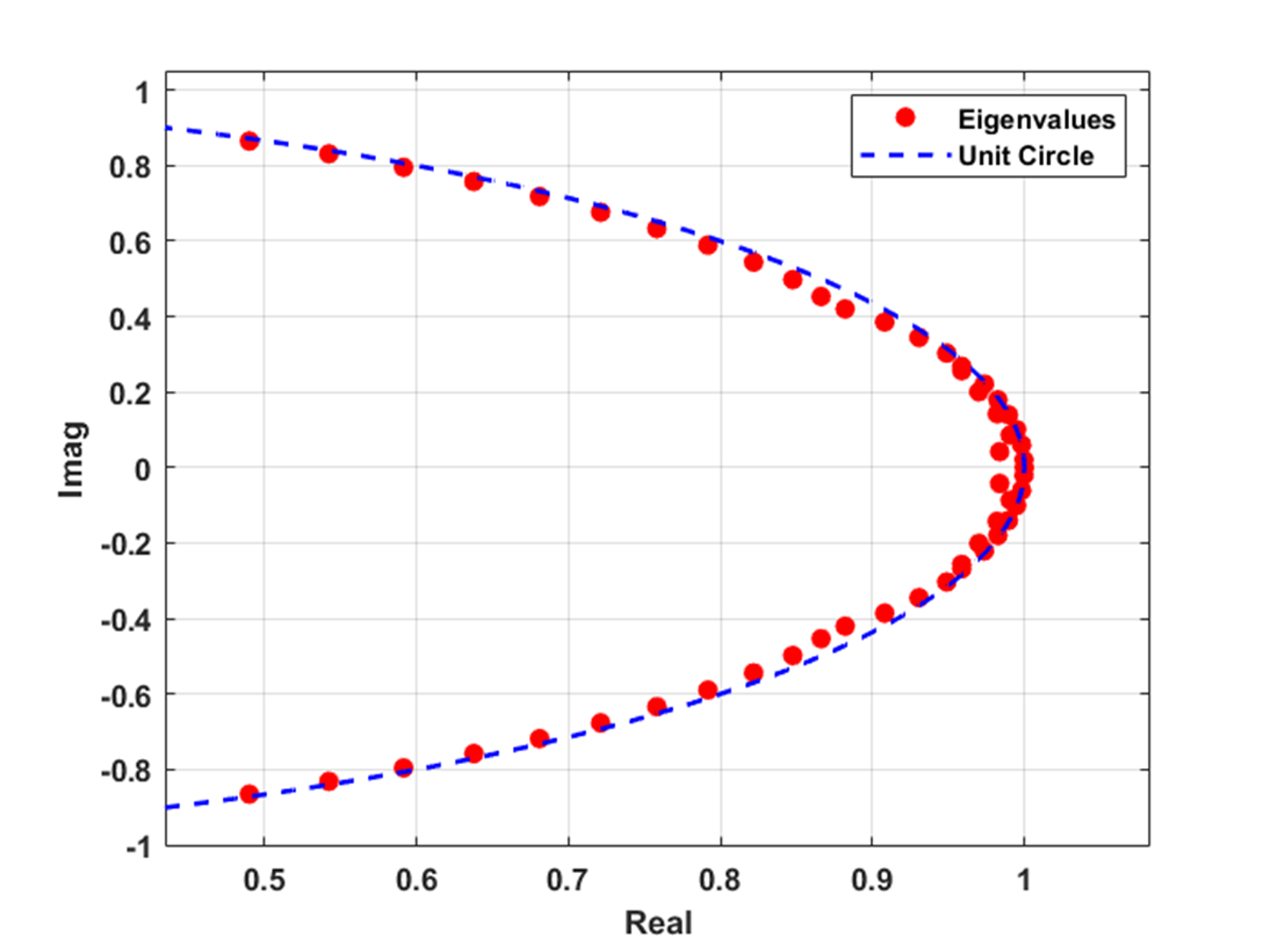}}
\end{minipage}
\caption{Eigenvalues of $\Tilde{\mbf{A}}$ matrix for each orbit type: (a) L1 Halo orbit (b) L1/L2 Butterfly orbit (c) Resonant (2:1) orbit}
\label{fig:Eigval}
\end{figure}

\begin{figure}[H]
\begin{minipage}{.48\linewidth}
\centering
\subfigure[]{\label{fig:L1HALO_plane:a}\includegraphics[scale=0.12]{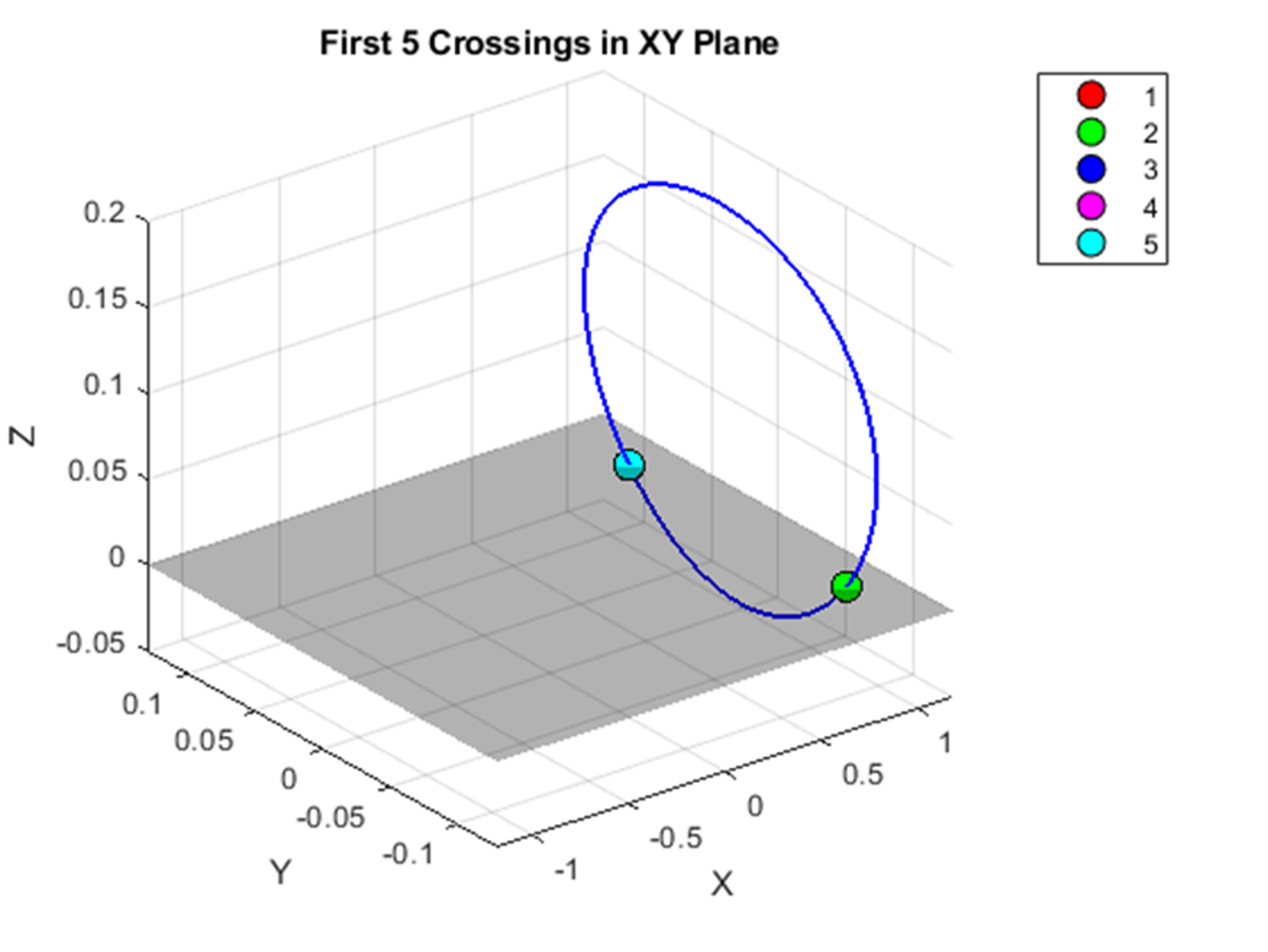}}
\end{minipage}%
\begin{minipage}{.48\linewidth}
\centering
\subfigure[]{\label{fig:L1HALO_plane:b}\includegraphics[scale=0.12]{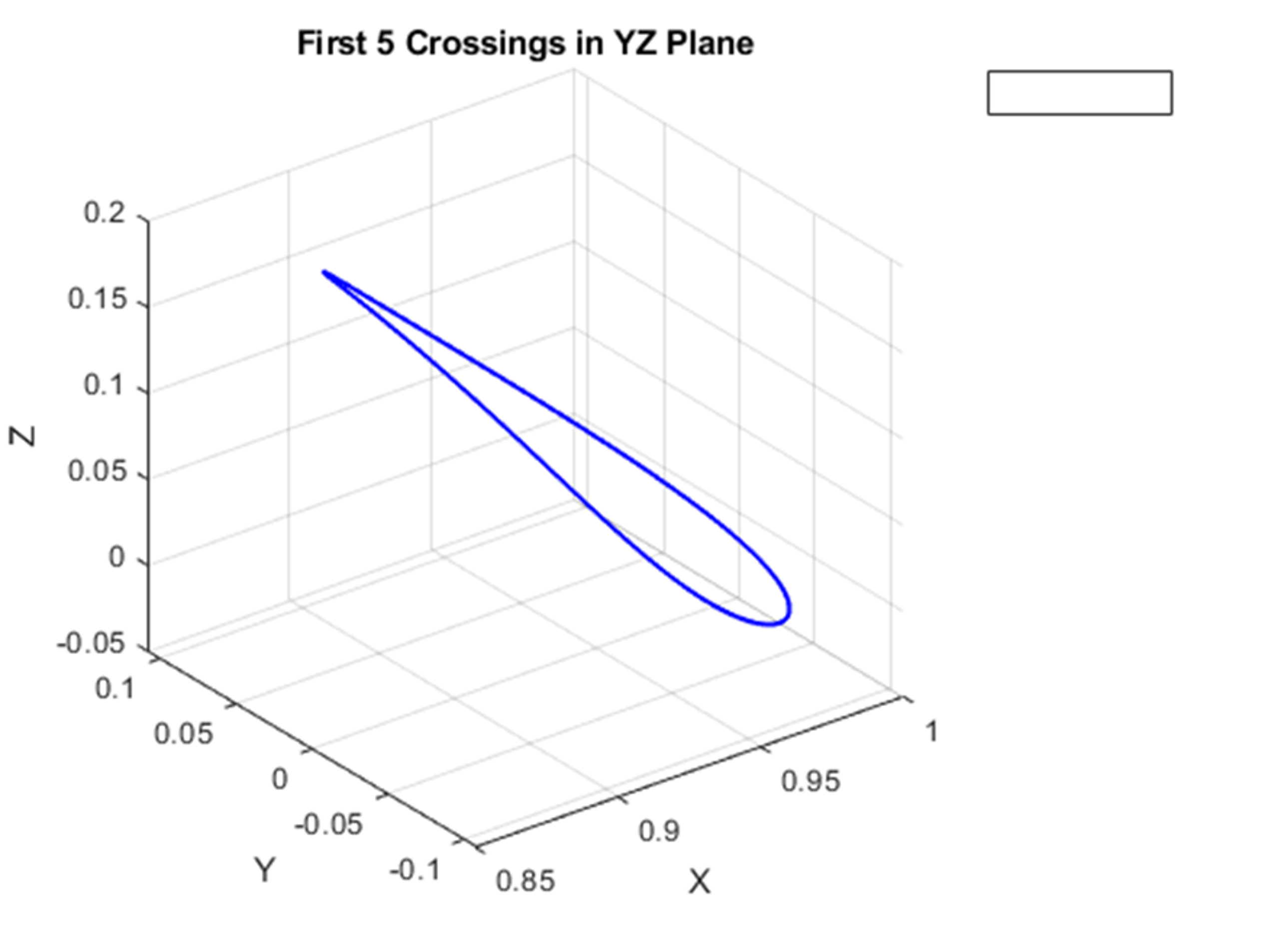}}
\end{minipage}
\centering
\begin{minipage}{.48\linewidth}
\centering
\subfigure[]{\label{fig:L1HALO_plane:c}\includegraphics[scale=0.12]{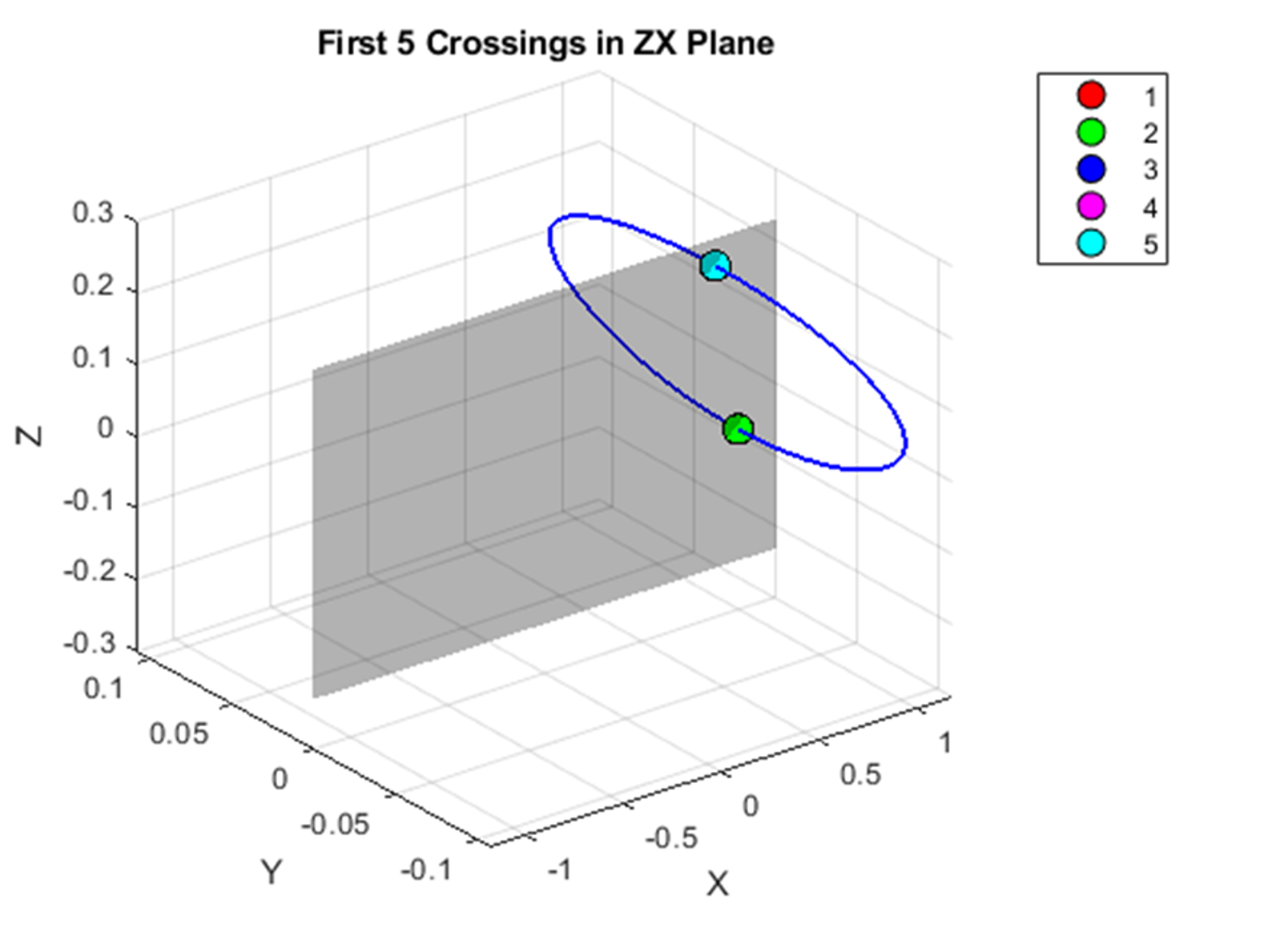}}
\end{minipage}
\caption{Initial Crossings of the L1 Halo Orbit with Coordinate Hyperplanes: (a) XY Plane, (b) YZ Plane, and (c) ZX Plane.}
\label{fig:L1HALO_plane}
\end{figure}

\begin{figure}[H]
\begin{minipage}{.48\linewidth}
\centering
\subfigure[]{\label{fig:L12BFLY_plane:a}\includegraphics[scale=0.12]{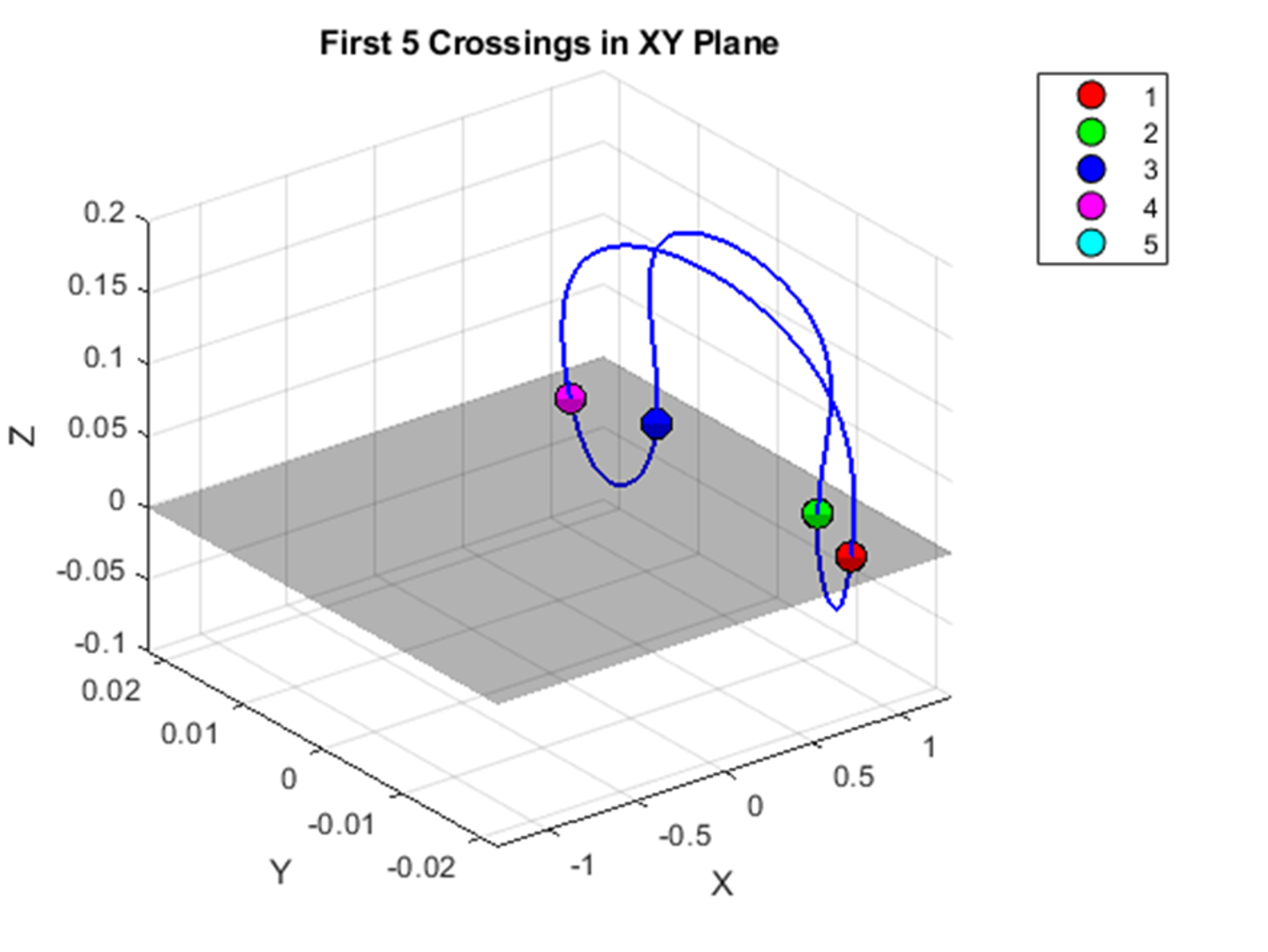}}
\end{minipage}%
\begin{minipage}{.48\linewidth}
\centering
\subfigure[]{\label{fig:L12BFLY_plane:b}\includegraphics[scale=0.12]{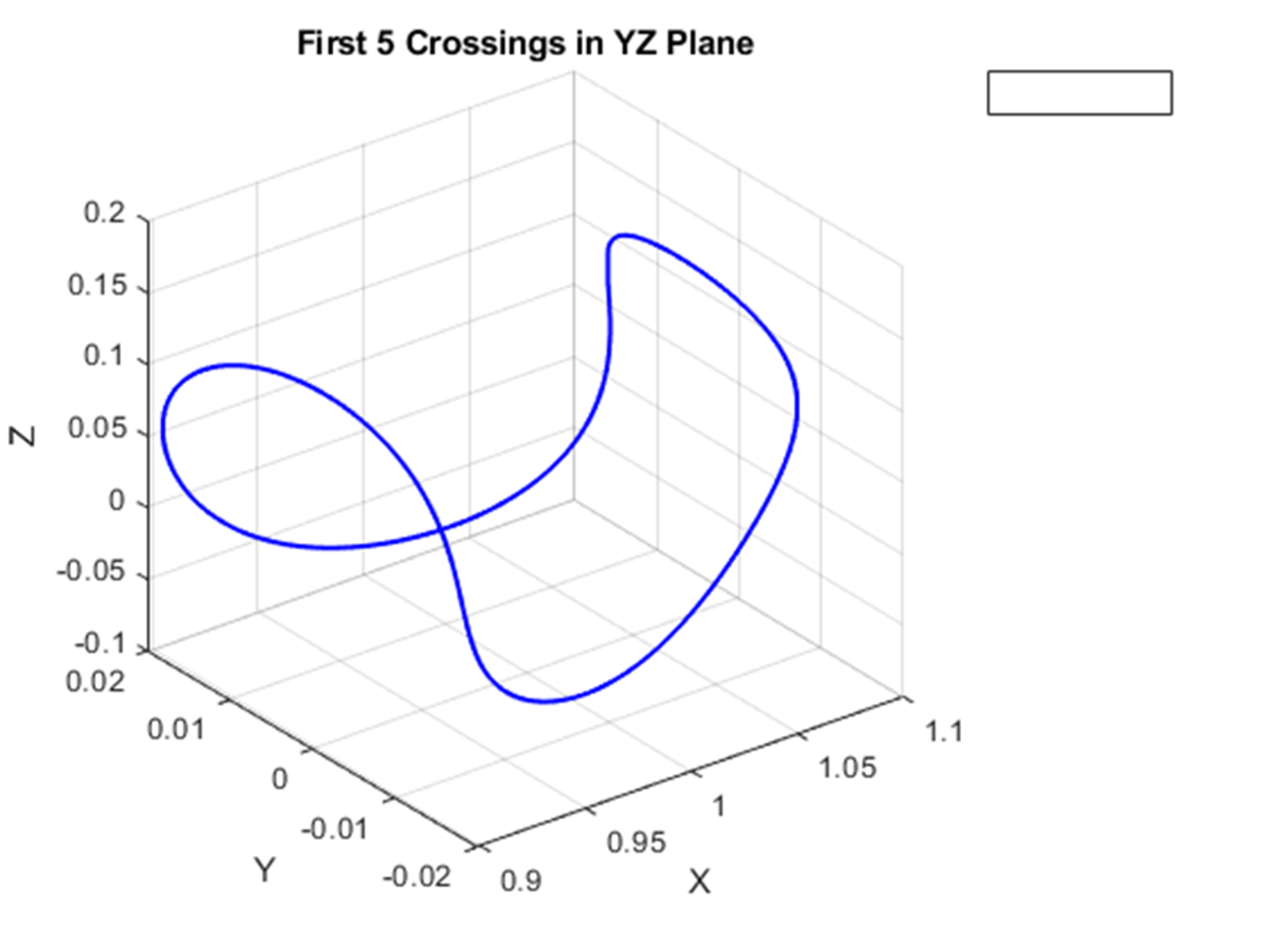}}
\end{minipage}
\centering
\begin{minipage}{.48\linewidth}
\centering
\subfigure[]{\label{fig:L12BFLY_plane:c}\includegraphics[scale=0.12]{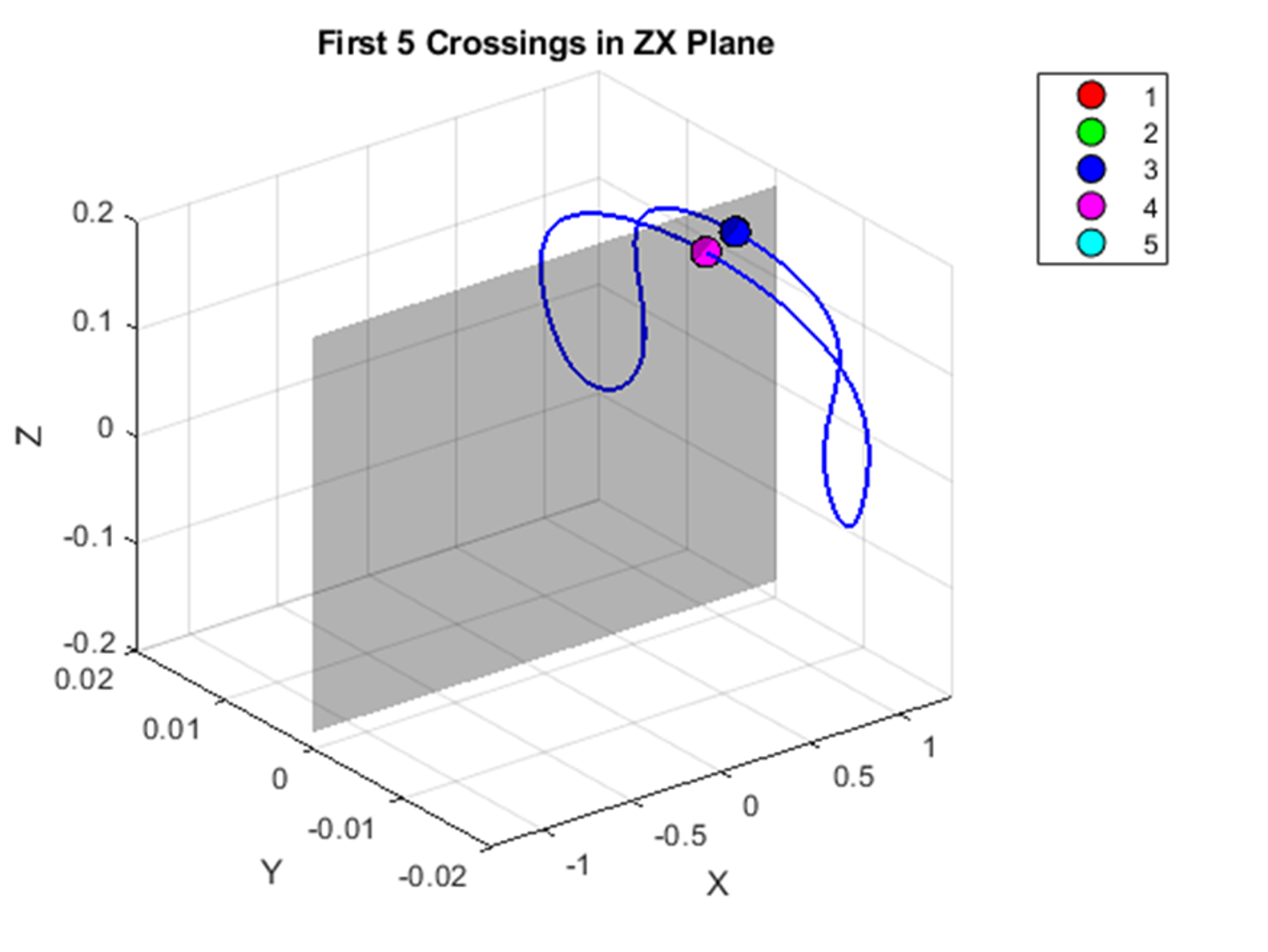}}
\end{minipage}
\caption{Initial Crossings of the L1/L2 Butterfly Orbit with Coordinate Hyperplanes: (a) XY Plane, (b) YZ Plane, and (c) ZX Plane.}
\label{fig:L12BFLY_plane}
\end{figure}

\begin{figure}[H]
\begin{minipage}{.49\linewidth}
\centering
\subfigure[]{\label{fig:Reso_plane:a}\includegraphics[scale=0.12]{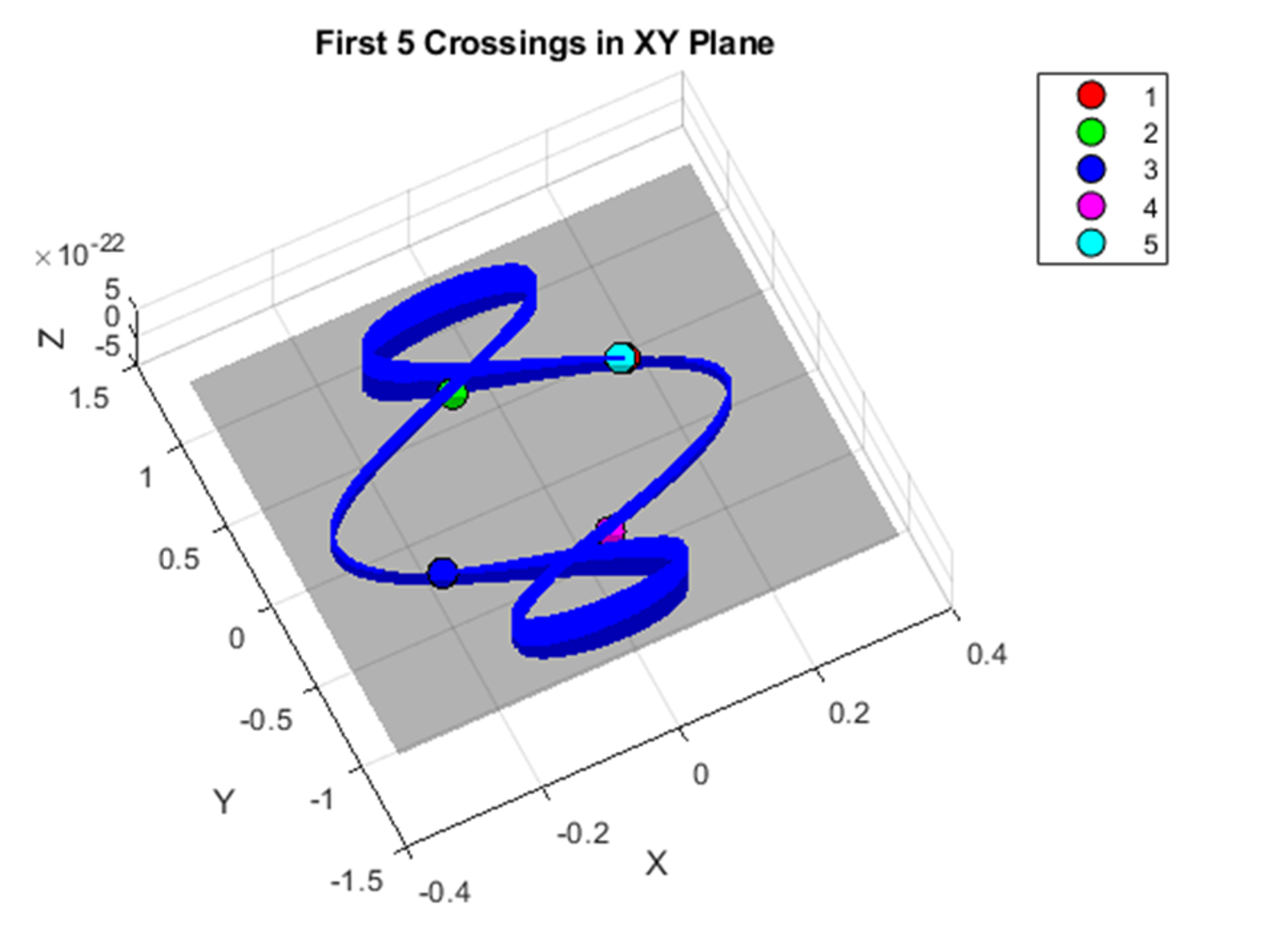}}
\end{minipage}%
\begin{minipage}{.49\linewidth}
\centering
\subfigure[]{\label{fig:Reso_plane:b}\includegraphics[scale=0.12]{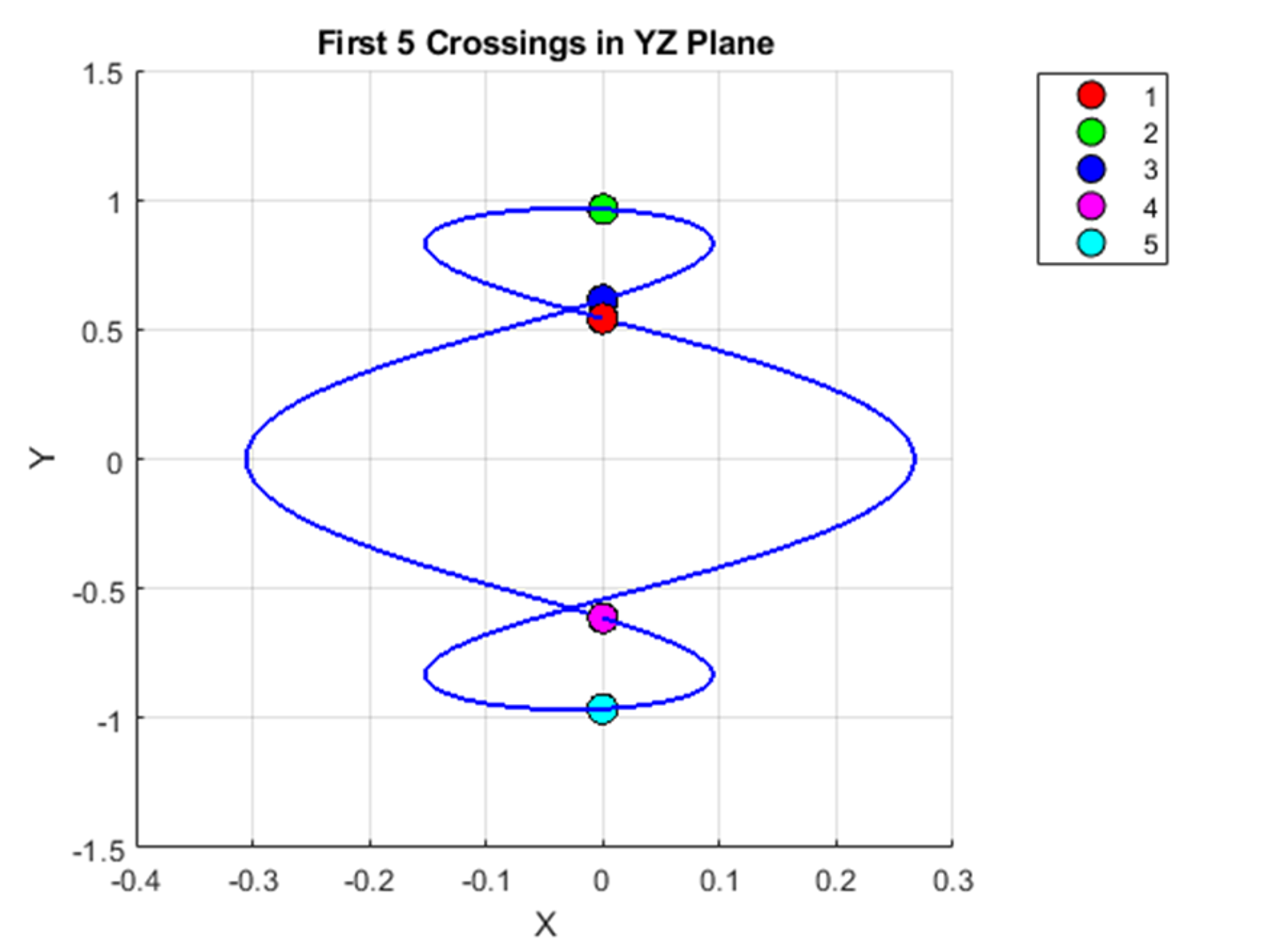}}
\end{minipage}
\centering
\begin{minipage}{.49\linewidth}
\centering
\subfigure[]{\label{fig:Reso_plane:c}\includegraphics[scale=0.12]{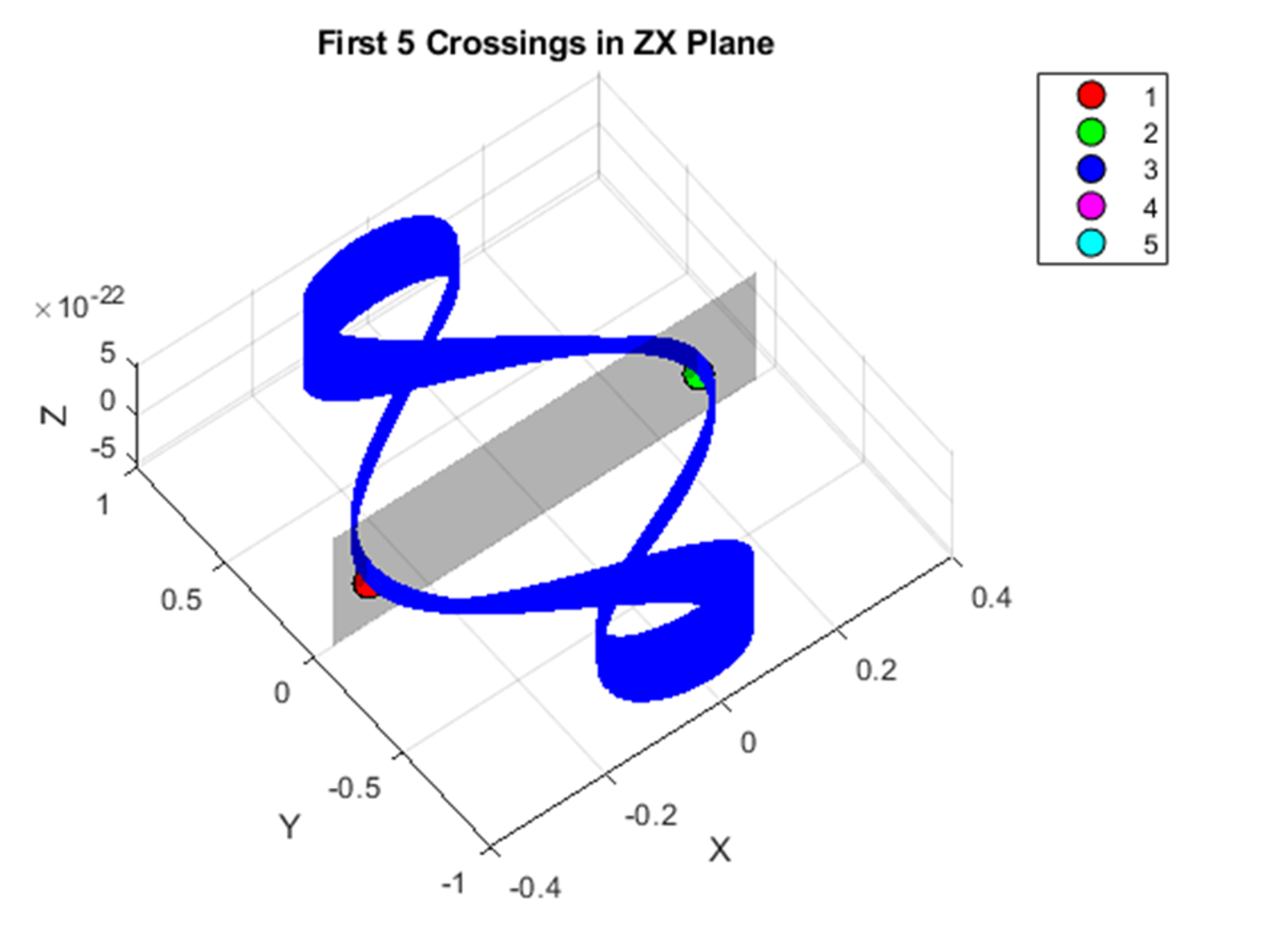}}
\end{minipage}
\caption{Initial Crossings of the Resonant (2:1) Orbit with Coordinate Hyperplanes: (a) XY Plane, (b) YZ Plane, and (c) ZX Plane.}
\label{fig:Reso_plane}
\end{figure}

Figure~\eqref{fig:SpecComp} presents the FFTs of both the original training dataset and the reconstructed DMD output over the same time interval for each scenario. These figures demonstrate that the DMD effectively captures the frequency content of the dataset. Additionally, this frequency plot can be used to determine the orbital period. In this work, the eigenvalues of the reduced-order matrix $\Tilde{\mbf{A}}$ are utilized. Specifically, the imaginary parts of the eigenvalues are employed to identify the orbit's frequencies, with the lowest frequency representing the fundamental frequency. The corresponding period can then be determined. Figure~\eqref{fig:Eigval} presents the eigenvalues of the $\Tilde{\mbf{A}}$ matrix for each case. In every scenario, the eigenvalues lie on or within the unit circle, indicating that the trajectories are stable.
To further validate the identified period, the time between successive Cartesian hyperplane crossings is counted. In this approach, the three coordinate hyperplanes are plotted alongside the 3D trajectory, allowing for the counting of crossings/intersections and measuring the duration between them to estimate the period. Figures~\eqref{fig:L1HALO_plane}, \eqref{fig:L12BFLY_plane}, and \eqref{fig:Reso_plane} illustrate this validation process for the three orbits analyzed in this work. This process is very similar to constructing Poincaré sections, with the primary objective here being the identification of the period of the closed orbit. Investigating dynamical properties such as identifying the Floquet matrix from the Poincare map and determining the stability of the fixed point will be considered in future work. Here, this process is purely used to validate the spectral content captured by DMD via estimates of the period.

\section{Conclusion \& Future Work}
\label{Sec:Conc}
This work examined the predictive capabilities of the DMD algorithm and its applicability in the cis-lunar orbital regime. A theoretical framework was provided for determining the minimum number of time delays required to model a nonlinear periodic trajectory using an autoregressive (AR) model, establishing connections to the Hankel-DMD algorithm. Additionally, methods for evaluating accuracy were detailed through comparisons of spectral content between the original training dataset and its DMD-reconstructed counterparts. Coordinate hyperplanes, analogous to Poincaré sections, were constructed to calculate the period of the orbit, thereby validating the periodic solution obtained by DMD.
A limitation of this work, however, is that the scenarios considered were restricted to closed orbits with stable trajectories. In future research, the authors plan to examine more unstable and chaotic trajectories that may be relevant to cislunar missions. This will involve leveraging the theory of time-delayed embedding and exploring the existence of strange attractors in such chaotic systems. Additionally, the use of Poincaré sections for such chaotic trajectories will be further investigated.

\section*{Acknowledgements}
The authors thank Dr. Michael Yakes for helping support this work via the Air Force Office of Scientific Research Grant No. FA9550-20-1-0083.

\bibliographystyle{AAS_publication}   
\bibliography{references}   

\end{document}